\documentclass[fleqn,10pt]{wlscirep}
\usepackage[utf8]{inputenc}
\usepackage[T1]{fontenc}
\usepackage{todonotes}
\usepackage{graphicx}
\usepackage{svg}
\usepackage{caption}
\usepackage{subcaption}
\usepackage{todonotes}
\usepackage{xspace}

\newcommand{\numElectrodes}{14\xspace}
\newcommand{\numParticipants}{16\xspace}
\newcommand{\bestHorizontalCorrelationEarEogRValue}{0.81\xspace}
\newcommand{\bestHorizontalCorrelationEarEogPValue}{0.01\xspace}
\newcommand{\bestHorizontalCorrelationCamRValue}{0.56\xspace}
\newcommand{\bestHorizontalCorrelationCamPValue}{0.02\xspace}
\newcommand{\bestHorizontalCorrelationChannelOne}{L8\xspace}
\newcommand{\bestHorizontalCorrelationChannelTwo}{R8\xspace}
\newcommand{\bestHorizontalCorrleationLeftEarEOGRValue}{0.52\xspace}
\newcommand{\bestHorizontalCorrleationLeftEarEOGPValue}{0.02\xspace}
\newcommand{\bestHorizontalCorrleationLeftEarCamRValue}{0.44\xspace}
\newcommand{\bestHorizontalCorrleationLeftEarCamPValue}{0.03\xspace}
\newcommand{\bestHorizontalCorrleationLeftEarChannelOne}{L1\xspace}
\newcommand{\bestHorizontalCorrleationLeftEarChannelTwo}{L4\xspace}
\newcommand{\bestHorizontalCorrleationRightEarEOGRValue}{0.41\xspace}
\newcommand{\bestHorizontalCorrleationRightEarEOGPValue}{0.03\xspace}
\newcommand{\bestHorizontalCorrleationRightEarCamRValue}{0.39\xspace}
\newcommand{\bestHorizontalCorrleationRightEarCamPValue}{0.04\xspace}
\newcommand{\bestHorizontalCorrleationRightEarChannelOne}{R5\xspace}
\newcommand{\bestHorizontalCorrleationRightEarChannelTwo}{R8\xspace}
\newcommand{\bestVerticalCorrelationEarEOGRValue}{0.28\xspace}
\newcommand{\bestVerticalCorrelationEarEOGPValue}{0.04\xspace}
\newcommand{\bestVerticalCorrelationCamRValue}{0.35\xspace}
\newcommand{\bestVerticalCorrelationCamPValue}{0.05\xspace}
\newcommand{\bestVerticalCorrelationChannelOne}{L1\xspace}
\newcommand{\bestVerticalCorrelationChannelTwo}{L8\xspace}

\newcommand{\bestVerticalCorrleationLeftEarChannelOne}{L1\xspace}
\newcommand{\bestVerticalCorrleationLeftEarChannelTwo}{L8\xspace}

\newcommand{\bestVerticalCorrleationRightEarChannelOne}{R2\xspace}
\newcommand{\bestVerticalCorrleationRightEarChannelTwo}{R7\xspace}
\newcommand{\saccadesleftRValue}{0.99\xspace}
\newcommand{\saccadesleftPValue}{0.0\xspace}
\newcommand{\voltageRValueleft}{0.99\xspace}
\newcommand{\voltagePValueleft}{0.0\xspace}

\newcommand{\saccadeVoltageDeflectionEarEOGleftSevenFive}{-14.16\xspace}

\newcommand{\saccadeVoltageDeflectionEarEOGleftOneFiveZero}{-39.21\xspace}

\newcommand{\saccadesrightRValue}{0.99\xspace}
\newcommand{\saccadesrightPValue}{0.0\xspace}
\newcommand{\voltageRValueright}{0.99\xspace}
\newcommand{\voltagePValueright}{0.0\xspace}

\newcommand{\saccadeVoltageDeflectionEarEOGrightSevenFive}{15.37\xspace}

\newcommand{\saccadeVoltageDeflectionEarEOGrightOneFiveZero}{40.16\xspace}

\newcommand{\saccadesupRValue}{0.6\xspace}
\newcommand{\saccadesupPValue}{0.23\xspace}
\newcommand{\voltageRValueup}{0.6\xspace}
\newcommand{\voltagePValueup}{0.23\xspace}

\newcommand{\saccadeVoltageDeflectionEarEOGupSevenFive}{-1.3\xspace}

\newcommand{\saccadeVoltageDeflectionEarEOGupOneFiveZero}{0.02\xspace}

\newcommand{\saccadesdownRValue}{0.19\xspace}
\newcommand{\saccadesdownPValue}{0.73\xspace}
\newcommand{\voltageRValuedown}{0.19\xspace}
\newcommand{\voltagePValuedown}{0.73\xspace}

\newcommand{\saccadeVoltageDeflectionEarEOGdownFiveZero}{-2.57\xspace}

\newcommand{\saccadeVoltageDeflectionEarEOGdownOneFiveZero}{-2.93\xspace}

\newcommand{\friedmanHorizontal}{0.0\xspace}
\newcommand{\friedmanVertical}{0.06\xspace}
\newcommand{\friedmanHorizontalTobii}{0.0\xspace}
\newcommand{\friedmanVerticalTobii}{0.68\xspace}
\newcommand{\numSaccadesLeftRight}{576\xspace}
\newcommand{\numValidSaccades}{565\xspace}

\newcommand{\linearRegressionMeanAbsErrorEarleft}{4.24\xspace}
\newcommand{\linearRegressionMeanAbsErrorGoldleft}{2.72\xspace}
\newcommand{\linearRegressionStdAbsErrorEarleft}{4.22\xspace}
\newcommand{\linearRegressionStdAbsErrorGoldleft}{2.29\xspace}
\newcommand{\linearRegressionMeanAbsErrorEarright}{4.45\xspace}
\newcommand{\linearRegressionMeanAbsErrorGoldright}{2.64\xspace}
\newcommand{\linearRegressionStdAbsErrorEarright}{3.74\xspace}
\newcommand{\linearRegressionStdAbsErrorGoldright}{2.15\xspace}
\newcommand{\linearRegressionMeanAbsErrorEar}{4.34\xspace}
\newcommand{\linearRegressionMeanAbsErrorGold}{2.68\xspace}
\newcommand{\linearRegressionStdAbsErrorEar}{3.99\xspace}
\newcommand{\linearRegressionStdAbsErrorGold}{2.23\xspace}
\newcommand{\numParticipantsMale}{11\xspace}
\newcommand{\numParticpantsFemale}{5\xspace}
\newcommand{\meanParticipantAge}{24.69\xspace}
\newcommand{\meanParticipantAgeSD}{5.25\xspace}
\newcommand{\numParticipantsEuropean}{12\xspace}
\newcommand{\numParticipantsAsian}{3\xspace}
\newcommand{\numParticipantsLatino}{1\xspace}
\newcommand{\numParticipantsBeardYes}{2\xspace}
\newcommand{\numParticipantsBeardNo}{12\xspace}
\newcommand{\numPariticpansBeardBarely}{2\xspace}
\newcommand{\numParticipantsHairLong}{6\xspace}
\newcommand{\numParticipantsHairShort}{8\xspace}
\newcommand{\numParticipantsHairMedium}{2\xspace}
\newcommand{\meanInsideEyeEyeDistance}{3.16\xspace}
\newcommand{\meanInsideEyeEyeDistanceSD}{0.54\xspace}
\newcommand{\meanOutsideEyeEyeDistance}{10.53\xspace}
\newcommand{\meanOutsideEyeEyeDistanceSD}{0.5\xspace}
\newcommand{\DistanceLeftEyeElectrodeCenter}{5.22\xspace}
\newcommand{\DistanceLeftEyeElectrodeCenterSD}{0.71\xspace}
\newcommand{\DistanceRightEyeElectrodeCenter}{5.22\xspace}
\newcommand{\DistanceRightEyeElectrodeCenterSD}{0.71\xspace}
\newcommand{\minParticipantAge}{20\xspace}
\newcommand{\maxParticipantAge}{38\xspace}

\definecolor{red}{RGB}{0,0,0}

\title{earEOG via Periauricular Electrodes to Facilitate Eye Tracking in a Natural Headphone Form Factor}

\author[1,*,**]{Tobias King}
\author[2,**]{Michael Knierim}
\author[1]{Philipp Lepold}
\author[3]{Christopher Clarke}
\author[4,5]{Hans Gellersen}
\author[1]{Michael Beigl}
\author[1]{Tobias Röddiger}

\affil[1]{Karlsruhe Institute of Technology, TECO / Pervasive Computing Systems, Karlsruhe, 76131, Germany}
\affil[2]{Karlsruhe Institute of Technology, Institute of Information Systems and Marketing, Karlsruhe, 76131, Germany}
\affil[3]{University of Bath, Human Computer Interaction, Bath, BA2 7AY, United Kingdom}
\affil[4]{Lancaster University, Interactive Systems, Lancaster, LA1 4YW, United Kingdom}
\affil[5]{Aarhus University, Department of Computer Science, Aarhus, 8200, Denmark}

\affil[*]{tobias.king@kit.edu}
\affil[{**}]{both authors contributed equally to this work}

\begin{abstract}

Eye tracking technology is frequently utilized to diagnose eye and neurological disorders, assess sleep and fatigue, study human visual perception, and enable novel gaze-based interaction methods.
However, traditional eye tracking methodologies are constrained by bespoke hardware that is often cumbersome to wear, complex to apply, and demands substantial computational resources.
To overcome these limitations, we investigated the application of Electrooculography (EOG) eye tracking using \numElectrodes electrodes positioned around the ears, integrated into a custom-built headphone form factor device.
In a controlled laboratory experiment, \numParticipants participants tracked a series of on-screen stimuli designed to induce smooth pursuits and saccades.
Data analysis identified the optimal electrode pairs for tracking vertical and horizontal eye movements, benchmarked against gold-standard EOG and camera-based eye tracking.
The electrode montage closest to the eyes provided the best results for horizontal eye movements.
One-dimensional smooth pursuit eye movements measured via earEOG exhibited a high correlation with the gold-standard for horizontal 1D pursuits spanning 2.5° to 15° visual angle for the best performing electrode pair ($\rm r_{EOG} = \bestHorizontalCorrelationEarEogRValue, p=\bestHorizontalCorrelationEarEogPValue$; $\rm r_{CAM} = \bestHorizontalCorrelationCamRValue, p=\bestHorizontalCorrelationCamPValue$).
Vertical 1D smooth pursuits were only weakly correlated for the best performing pair ($\rm r_{EOG} = \bestVerticalCorrelationEarEOGRValue, p=\bestVerticalCorrelationEarEOGPValue$; $\rm r_{CAM} = \bestVerticalCorrelationCamRValue, p=\bestVerticalCorrelationCamPValue$).
Voltage deflections of earEOG and gold-standard EOG for saccades from 2.5° to 15° in the four cardinal directions are highly correlated for horizontal eye movement ($\rm r_{left} = \voltageRValueleft, p=\voltagePValueleft$; $\rm r_{right} = \voltageRValueright, p=\voltagePValueright$) but not for vertical eye movements ($\rm r_{up} = \voltageRValueup, p=\voltagePValueup$; $\rm r_{down} = \voltageRValuedown, p=\voltagePValuedown$).
A regression model was employed to predict absolute gaze angle changes of horizontal saccades using earEOG and gold-standard EOG.
In the left and right directions, the earEOG model achieved a mean absolute angular error of $\rm \linearRegressionMeanAbsErrorEar^\circ \pm \linearRegressionStdAbsErrorEar^\circ$, for saccades ranging from 2.5° to 15°.
In comparison, gold-standard EOG attained mean absolute angular error of $\rm \linearRegressionMeanAbsErrorGold ^\circ\pm \linearRegressionStdAbsErrorGold^\circ$.
Overall, horizontal earEOG demonstrated strong performance, indicating its potential effectiveness in our setup.
On the other hand, vertical earEOG showed significantly poorer results, suggesting that it may not be feasible with our current configuration.

\end{abstract}

\begin{document}

\flushbottom
\maketitle
\thispagestyle{empty}

\section*{Introduction}
Eye tracking is a widely employed technique for sensing and interaction that typically involves either camera-based\cite{gibaldi2017evaluation, chennamma2013survey} or Electrooculography (EOG) \cite{dhuliawala2016smooth, bulling2010eye, chennamma2013survey} methods. While camera-based eye tracking can be precise, it is computationally intensive and requires a significant amount of power \cite{chennamma2013survey}. On the other hand, traditional EOG is less computationally demanding and even works when the eyes are closed. However, it is restricted to tracking relative changes in gaze direction, is subject to signals drift and is relatively invasive as electrodes have to be glued on the face around the eyes. Despite these limitations, EOG has various interesting applications. In a medical context, EOG can be applied in sleep studies \cite{zhu2014eog} or to diagnose balance disorders \cite{goebel1992prevalence}. Outside the clinic, EOG can be used for hands-free interaction with wearable devices \cite{lee2016real, dhuliawala2016smooth, 10.1145/3715669.3723110}, to classify user activities\cite{bulling2010eye}, to quantify reading activity up to word-level accuracy \cite{bulling2008robust, kunze2015much}, and for providing directed auditory attention in noisy environments \cite{favre2017steering, favre2017real, favre2019absolute}.

To make EOG more feasible, past work implemented electrodes into smart glasses which improves wearability and more naturally integrates into the everyday-life of the user \cite{dhuliawala2016smooth, kunze2015much}.
Prior research has also suggested that eye tracking using electrodes placed inside the ear canal\cite{favre2017real, favre2019absolute, hladek2018real, manabe2013conductive}  or at the mandible\cite{manabe2006full} is generally feasible. 
Favre-Felix et al.\cite{favre2019absolute} investigated the use of ear-based EOG and motion sensors around the ear to estimate absolute horizontal eye gaze in multi-talker situations, showing promising results when the head was fixed. However, hardware issues hindered reliable estimations when the head was free. 
Manabe et al.\cite{manabe2013conductive} developed an earphone-based eye gesture input interface using conductive rubber electrodes. 
In another paper, the same authors\cite{manabe2006full} proposed a headphone-type gaze detector with electrodes placed at the mandible on one ear to estimate gaze direction. Based on experiments with a single user, they achieve an estimation error of 4.4° (horizontal) and 8.3° (vertical) in a 5~$\rm\times$~3 fixation point grid (20\textdegree~visual angle between fixation points) and lay the foundation for our work.
Similar to smart glasses, ear-based EOG devices have the potential to be more comfortable, discreet, and portable than traditional EOG.
Furthermore, the ear is an ideal location for integrating eye tracking with audio applications for example for directed auditory attention in noisy environments and with hearing aids\cite{favre2017real, favre2019absolute, hladek2018real}.

The effectiveness of in-ear-based eye tracking for horizontal eye movements was explored in related work \cite{favre2017real, favre2019absolute, hladek2018real, manabe2013conductive}. However, incorporating eye-tracking capabilities into headphones could have several advantages over the in-ear EOG method that we explore in this paper. 
Firstly, the proximity of headphones to the eyes enhances the sensitivity to eye movements, leading to potentially more accurate measurements of changes in eye position. 
To this extent, we seek to expand upon the initial work of Manabe et al.\cite{manabe2006full} and evaluate headphone-based eye tracking in-depth with a large number of participants.

\begin{figure}
    \centering
    \begin{subfigure}[t]{0.3\textwidth}
        \centering
        \caption{earEOG headphones.}
        \includegraphics[height=4cm, page=4]{figures/overleaf-crop.pdf}
        \label{fig:headphone_rendering}
    \end{subfigure}
    \begin{subfigure}[t]{0.45\textwidth}
        \centering
        \caption{Electrode positions.}
        \includegraphics[height=4cm]{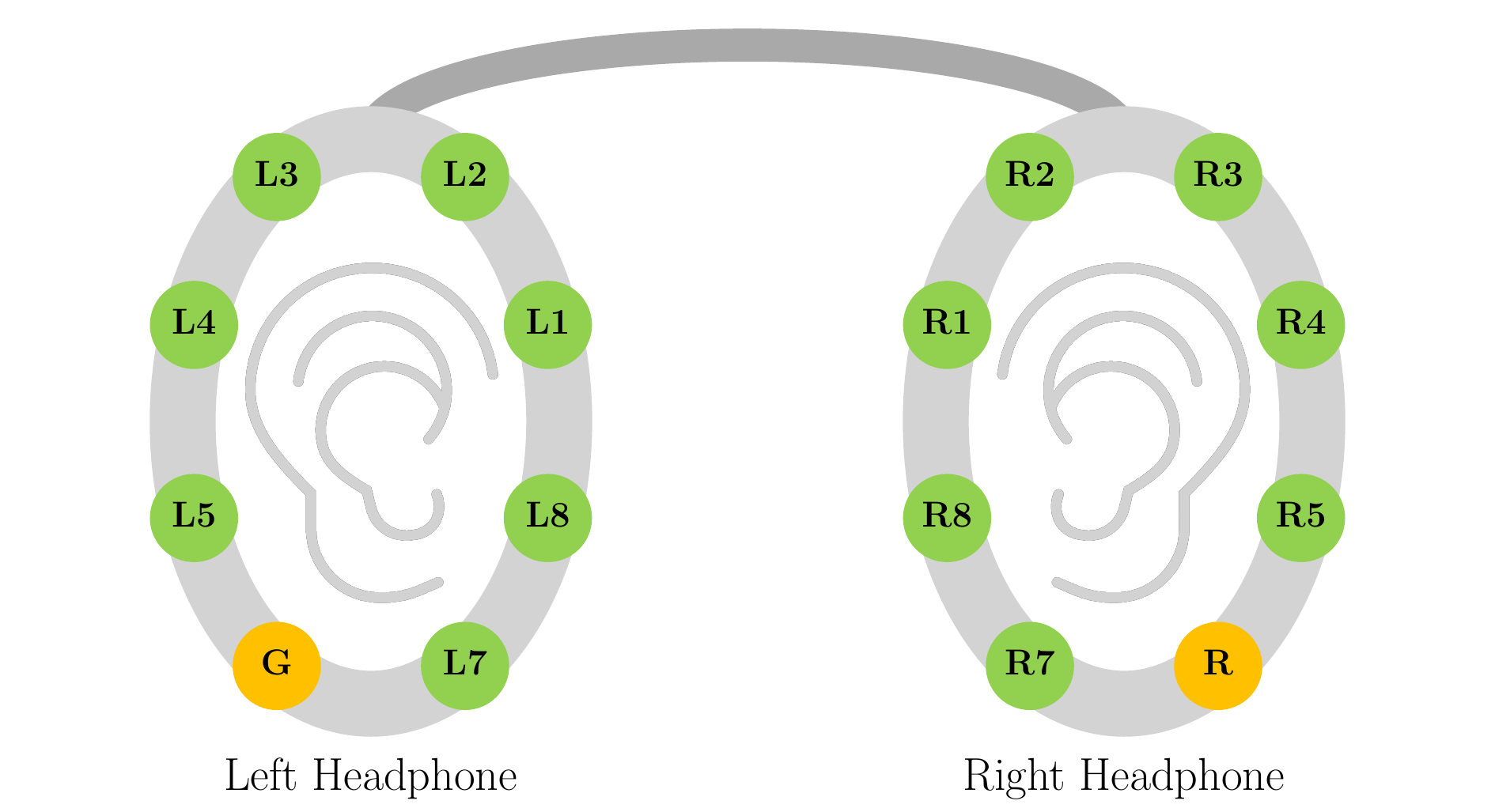}
        \label{fig:headphone_base_image}
    \end{subfigure}
    \begin{subfigure}[t]{0.2\textwidth}
        \centering
        \caption{Study participant.}
        \includegraphics[height=4cm]{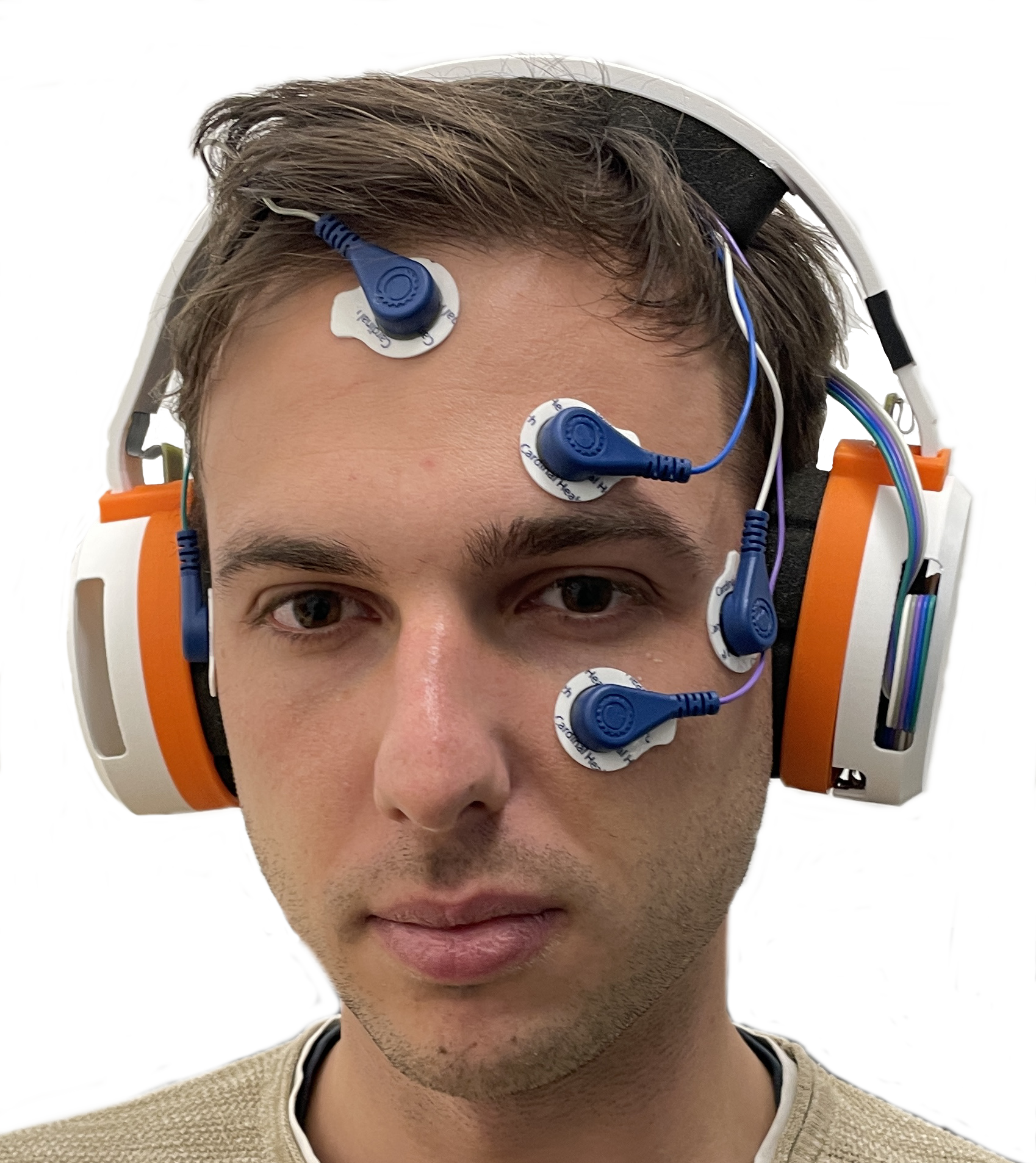}
        \label{fig:apparatus}
    \end{subfigure}
    \caption{\textbf{Overview of earEOG headphones showing the electrode positions and study apparatus.} \ref{fig:headphone_rendering} The EOG headphones have a standard headphone form factor with 16 gold-plated electrodes (copper base) on each ear (One reference (R) and one Ground (G) electrode). The left headphone includes the Cyton Board while the right headphone includes the battery;
    \ref{fig:headphone_base_image} Schematic drawing showing the positioning of the electrodes around the ears;
    \ref{fig:apparatus} A study participant wearing the EOG headphones and gold-standard Electrooculography simultaneously.}
    \label{fig:overview}
\end{figure}

We thoroughly investigate the hypothesis that EOG-based eye tracking using electrodes placed around the ears in a regular headphone form factor is a reliable and accurate method for studying eye movements.
To understand the achievable performance and add context, we ground our research in comparison to gold-standard EOG and camera-based eye tracking data.
For our evaluation, a specialized headphone device was developed with \numElectrodes electrodes positioned strategically around the ears, see \autoref{fig:overview}. 
Using the earEOG headphones, we conducted a lab study with \numParticipants participants to collect data of two tasks.

For the first task, participants were asked to follow one-dimensional moving targets to elicit smooth pursuit eye motions\cite{vidal2013pursuits}.
Smooth pursuits are continuous eye movements that allow the eyes to follow a moving target smoothly without any jerks or abrupt changes in direction\cite{robinson1965mechanics}. 
Smooth pursuits provide a continuous signal that can be more easily correlated with the changes in eye position than abrupt saccades, enabling a better analysis of the relationship between electrode signals and eye movements. This, in turn, helps us to determine the most effective electrode positions for capturing the nuances of eye movement and to evaluate the overall feasibility of the earEOG method.
For horizontal eye movements, the bipolar montage of \bestHorizontalCorrelationChannelOne-\bestHorizontalCorrelationChannelTwo (difference in electrical signals between the two electrodes) yielded the highest correlation to horizontal gold-standard EOG ($\rm r_{EOG-\bestHorizontalCorrelationChannelOne-\bestHorizontalCorrelationChannelTwo} = \bestHorizontalCorrelationEarEogRValue, p=\bestHorizontalCorrelationEarEogPValue$) and also camera-based eye tracking ($\rm r_{CAM-\bestHorizontalCorrelationChannelOne-\bestHorizontalCorrelationChannelTwo} = \bestHorizontalCorrelationCamRValue, p=\bestHorizontalCorrelationCamPValue$). 
Using electrodes from just one ear, the montage of \bestHorizontalCorrleationLeftEarChannelOne-\bestHorizontalCorrleationLeftEarChannelTwo ($\rm r_{EOG-\bestHorizontalCorrleationLeftEarChannelOne-\bestHorizontalCorrleationLeftEarChannelTwo}=\bestHorizontalCorrleationLeftEarEOGRValue, p=\bestHorizontalCorrleationLeftEarEOGPValue$; $\rm r_{CAM-\bestHorizontalCorrleationLeftEarChannelOne-\bestHorizontalCorrleationLeftEarChannelTwo} = \bestHorizontalCorrleationLeftEarCamRValue, p=\bestHorizontalCorrleationLeftEarCamPValue$) and \bestHorizontalCorrleationRightEarChannelOne-\bestHorizontalCorrleationRightEarChannelTwo ($\rm r_{EOG-\bestHorizontalCorrleationRightEarChannelOne-\bestHorizontalCorrleationRightEarChannelTwo}=\bestHorizontalCorrleationRightEarEOGRValue, p=\bestHorizontalCorrleationRightEarEOGPValue$; $\rm r_{CAM-\bestHorizontalCorrleationRightEarChannelOne-\bestHorizontalCorrleationRightEarChannelTwo} = \bestHorizontalCorrleationRightEarCamRValue, p=\bestHorizontalCorrleationRightEarCamPValue$) produced the highest correlation.
For vertical eye movements, only the montage of \bestVerticalCorrelationChannelOne-\bestVerticalCorrelationChannelTwo had significant but very weak correlation ($\rm r_{EOG-\bestVerticalCorrelationChannelOne-\bestVerticalCorrelationChannelTwo}=\bestVerticalCorrelationEarEOGRValue, p=\bestVerticalCorrelationEarEOGPValue$; $\rm r_{CAM-\bestVerticalCorrelationChannelOne-\bestVerticalCorrelationChannelTwo}=\bestVerticalCorrelationCamRValue, p=\bestVerticalCorrelationCamPValue$). The higher correlation measured on the left ear is likely due to the gold-standard EOG being attached to the left eye.

For the second task, participants were instructed to follow a point that jumped from the center of the screen at 0\textdegree~to 2.5\textdegree~up to 15\textdegree~in the four cardinal directions at 2.5\textdegree~increments.
Using the ideal electrode positions identified from the previous analysis it was found that voltage deflections during saccades of earEOG and gold-standard EOG across all angles are mostly highly correlated for horizontal saccades ($\rm r_{left}=\saccadesleftRValue, p=\saccadesleftPValue$; $\rm  r_{right}=\saccadesrightRValue, p=\saccadesrightPValue$). 
Vertical saccades were not significantly correlated to gold-standard EOG voltage deflections ($\rm r_{up}=\saccadesupRValue, p=\saccadesupPValue$; $\rm r_{down}=\saccadesdownRValue, p=\saccadesdownPValue$). 

Building upon the relationship between voltage deflections and the underlying gaze angle, a regressions model was evaluated to calculate the absolute saccade angle from earEOG for the horizontal direction. 
On average, horizontal earEOG achieved an absolute angular accuracy of $\linearRegressionMeanAbsErrorEar$°~$\rm \pm~\linearRegressionStdAbsErrorEar$°. In comparison, a similar model fitted on gold-standard horizontal EOG data achieves an absolute angular accuracy of $\rm \linearRegressionMeanAbsErrorGold$°~$\rm \pm~\linearRegressionStdAbsErrorGold$°.
These findings demonstrate that earEOG is a promising method for eye tracking on the horizontal axis, showing good correlation with gold-standard EOG, which indicates its potential usability. Moreover, it could be a good approach for easily integrable, user-friendly electrodes in everyday life.

In sum, our contributions are: (i) a thorough investigation of EOG-based eye tracking using electrodes placed around the ears in a custom-built headphone form factor, providing earEOG - a novel approach to on-the-go wearable eye tracking in headphones; (ii) an evaluation of different electrode positions for earEOG and recommendations for the optimal placement of electrodes, to enable a more effective use of earEOG for eye tracking, thereby enabling a wider range of applications; and
(iii) an evaluation of earEOG to predict absolute gaze angles in comparison to gold-standard EOG;

\section*{Methods}
\subsection*{Participants}
The participants in this study were recruited through a sample of convenience. Demographic information was self-reported by participants using questionnaires. 
Participants without sufficient vision capabilities, as determined by the need to wear glasses, were excluded from the study. The study was conducted in a controlled environment to minimize the influence of external factors (45\% humidity, 21°C room temperature).

The study protocol was approved by the institutional review board of the Karlsruhe Institute of Technology (Germany) and followed all relevant ethical regulations in accordance with the Declaration of Helsinki. Before beginning the study, participants were informed of the study protocol, the purpose of data collection, and the specific data that would be collected. Informed consent was obtained from participants through the signing of an agreement form. Participants were not compensated.

\subsection*{Apparatus}
A \numElectrodes-channel electric potential data collection setup was implemented using an OpenBCI Cyton + Daisy biosensing board, with 7 channels around each ear.
Additionally, the device included one ground electrode and one reference electrode.
Vertical and horizontal gold-standard EOG was also acquired using the OpenBCI board via four standard Ag/AgCl gel sticky electrodes glued around the eyes, see \autoref{fig:overview}~(a)~to~(c) (Informed consent to publish the identifiable images was obtained from all the participants). The OpenBCI was housed in a 3D-printed headphone enclosure. 
The gold-plated (copper base) electrodes around each ear are custom-built flex-PCBs that were fit to the headphone form factor.
Such electrodes are also common in exisiting wearable EEG headsets (e.g., Open-cEEGrid \cite{OpencEEG55:online}).
Before application, the area around the ears was cleaned with isopropyl alcohol. 
All EOG data was sampled at 125~Hz.
In addition, a stationary camera-based eye tracker (Tobii eyeX Pro) was employed to record ground-truth gaze angles at 60 Hz.

For data collection, a web-based tool was implemented that provides instructions to the participants and shows on-screen stimuli to elicit specific eye movement patterns.
The experiment was conducted using a 23" monitor (1920px $\times$ 1080px) with a viewing distance of 50 cm. Participants were seated centered in front of the screen, with the vertical center of the screen positioned at eye level using a height-adjustable desk. This ensured that the visual stimuli could be presented at the desired size and angle in relation to the participants head and the screen size.
The participant's head position was not fixed in space, but the person conducting the study carefully monitored the head position and distance to the screen to intervene if participants had moved significantly during data collection.

\paragraph{Data Collection Procedure}
After arriving at the lab, signing the consent forms, and asking any possible questions, participants were fitted with the earEOG headphones as well as gold-standard vertical and horizontal EOG electrodes.
They were then seated in front of the screen.
The stationary eye tracker was calibrated using the built-in 9-point calibration procedure.
Participants then followed to on-screen instructions to complete the eye-based tasks for data collection.
All tasks took approximately 7 minutes to complete and were repeated three times, totaling approximately 25 minutes of experiment duration per participant. Between each task, participants had a 10 seconds resting period and between each cycle, participants could rest their eyes freely for one minute.
The order of tasks was not counterbalanced.

\subsection*{Task, Stimuli, and Procedure}
Participants were presented with three different tasks to collect two types of eye movements: smooth pursuits and saccades. Smooth pursuits were added to find the best electrode for vertical and horizontal eye movement tracking, respectively. Saccades were added as they are fundamental for many research studies, and the absolute angle is the most characteristic to understand eye movements.

\paragraph{Smooth Pursuit Task}
Smooth pursuit eye movements are continuous eye movements that follow a moving object, and they exhibit much slower characteristics compared to rapid eye movements such as saccades that shift the gaze from one object or point in space to another. 
The smooth pursuit task consisted of 1D smooth pursuits angles in both vertical and horizontal directions. Participants were instructed to follow the gaze target that moved within a 2.5 to 15 degrees visual opening angle from the center for 6 seconds each. The frequency of the gaze target movement was set to 0.33, 0.5, and 1 Hz. The motion of the gaze target was eased by sin(x) on the axis of movement, which means it followed a simple harmonic motion. The gaze target had a diameter of 30 pixels, which is equivalent to 7.8 mm.
The purpose of this task was to find the best electrode for vertical and horizontal eye movement tracking. The eye movement data collected during the smooth pursuit task will be correlated with the eye movements of the gold-standard EOG and camera-based gaze signals. 

\begin{figure}
    \centering
    \begin{subfigure}[t]{0.19\textwidth}
        \centering
        \caption{Smooth pursuit angles.}
        \includegraphics[height=3cm, page=5]{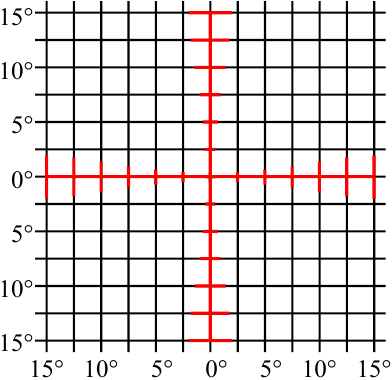}
        \caption*{
            \centering
            \parbox{0.9\linewidth}{
                Start position is set randomly within angle.
            }
        }
        \label{fig:pursuit-va}
    \end{subfigure}
    \hfill
    \begin{subfigure}[t]{0.38\textwidth}
        \centering
        \caption{Smooth pursuit visual targets.}
        \includegraphics[height=3cm, page=2]{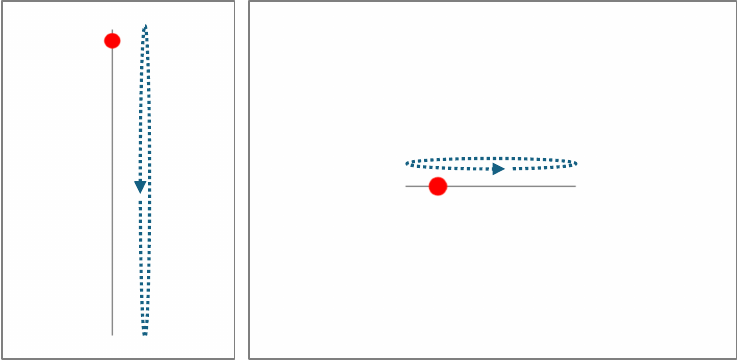}
        \caption*{
            \centering
            \parbox{0.9\linewidth}{
                Dot moves on horizontal or vertical line with simple harmonic motion.
            }
        }
        \label{fig:pursuit-target}
    \end{subfigure}
    \hfill
    \begin{subfigure}[t]{0.19\textwidth}
        \centering
        \caption{Saccade Angles.}
        \includegraphics[height=3cm, page=3]{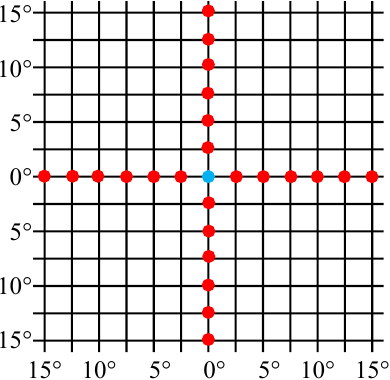}
        \caption*{
            \centering
            \parbox{0.9\linewidth}{
                Blue dot is the center resting point.
            }
        }
        \label{fig:saccade-angles}
    \end{subfigure}
    \hfill
    \begin{subfigure}[t]{0.2\textwidth}
        \centering
        \caption{Saccades visual targets.}
        \includegraphics[height=3cm, page=4]{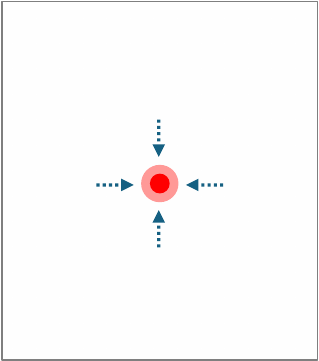}
        \caption*{
            \centering
            \parbox{0.9\linewidth}{
                Dot shrinks during fixation.
            }
        }
        \label{fig:saccade-target}
    \end{subfigure}
    
    \caption{\textbf{Overview of the different gaze stimuli and tasks presented to study participants.}   \ref{fig:pursuit-va} 1D smooth pursuit points in horizontal and vertical directions with opening angles of 2.5° to 15°;
    \ref{fig:pursuit-target} visual target moving on a straight line with simple harmonic motion that is shown to participants to elicit smooth pursuit eye motion; \ref{fig:saccade-angles} fixations points in four cardinal directions with 2.5° increments and center resting point; \ref{fig:saccade-target} visual target point shown to participants that shrinks during fixation.}
    \label{fig:stimuli}
\end{figure}

\paragraph{Saccade Task}
In the saccade task, participants were presented with a series of fixation points to elicit saccade eye motion. Participants started with resting their gaze in the resting position at the screen's center (0° visual angle in the x and y direction). The fixation point then jumped to 2.5-degree increments in each of the four cardinal directions (left, right, up, down) from 2.5 to 15 degrees. Each fixation point was presented for 2 seconds and shrank from 30 to 20 pixels (equivalent to 7.8 to 5.2~mm). After completing an angle, the fixation point returned to the center resting point for 2 seconds.
The purpose of the saccade task was to collect saccade eye movement data to understand the relationship between earEOG signal deflections, gold-standard EOG, and absolute gaze angle changes.

\subsection*{Data Analysis}
\subsubsection*{Preprocessing}
The collected gold-standard EOG and earEOG data per ear are already aligned as both were sampled using the same device.
For both the gold-standard EOG and the earEOG data, the signals for each channel were bandpass filtered between 0.1 and 15 Hz using a 5th-order Butterworth filter to eliminate noise and other artifacts, similar to methodologies employed in previous studies \cite{awsmabdullahReviewFilteringTechniques2023, s23062944}.
These pre-processing steps were carried out to ensure that the data was suitable for subsequent analysis.

\subsubsection*{Smooth Pursuit Analysis}
In order to find the best electrode positions for measuring eye movements with electric potentials from around the ears, we used the ground truth gold-standard EOG as well as the camera-based eye tracker data and correlated them with differential earEOG electrode signals.
We utilized the 1D smooth pursuit data (vertical and horizontal) to compute the best electrodes.
Smooth pursuit eye movements were chosen because they exhibit continuous electric signal changes compared to rapid saccades, which create a sharp, short spike in the EOG signal. Consequently, smooth pursuits are less prone to artefacts and noise and allow for a continuous sequence to be correlated between the gold-standard and ear-based EOG principle, which improves the reliability of our results.

For horizontal eye movements, we considered electrode combinations which were aligned with the horizontal axis of the head (see \autoref{fig:corr_horizontal}).
Similarly, for vertical eye movements, we analyzed a subset of electrode pairs, specifically focusing on those positioned vertically above one another (see \autoref{fig:corr_vertical}).
Diagonal electrode combinations were not considered, as they measure both horizontal and vertical components simultaneously, leading to ambiguities.
For all recorded smooth pursuits, the entire six-second sample was used for correlation.
To account for signal propagation delays, we permitted a maximum lag of 12 samples ($\approx 100\,\mathrm{ms}$) when calculating correlations between EOG channels.
We also allowed up to 64 samples ($\approx 500\,\mathrm{ms}$) for correlations between EOG and the eye tracker as the Tobii eye tracker and the EOG data were not perfectly time-aligned.
To correlate the data, we first preprocess the EOG and eye tracker data.
For each saccade in the EOG data, we detrend the singal, apply a mean filter with filter length 50 and normalize the data to be in the interval [-1, 1].
For each saccade as recorded by the eye tracker, we interpolate its values as some may be missing.
Then, we resample the data to match the sampling rate of the EOG data (125 Hz), apply a butterworth bandpass filter of order 5 between 0.1 Hz and 15 Hz, apply a mean filter with filter length 50 and finally normalize the data to be in the interval [-1, 1].
To compute the mean over the correlation coefficients, we first apply the Fisher z-transformation.
The results are based on all sampled smooth pursuit speeds and angles.
To identify the best vertical electrodes, we only used the vertical smooth pursuit data, and for the horizontal electrodes, we only use horizontal smooth pursuit data.

To identify whether the correlations between electrode montages and the gold-standard differ significantly, we employ a Friedman test.
We compare each direction (horizontal and vertical) separately and use the gold-standard EOG as well as the camera-based eye tracker data as ground truth.\\
If the p-value of the Friedman test is below 0.05, we conduct a Wilcoxon signed-rank test (p-values are Bonferroni-corrected) to assess pairwise differences in the correlation values of each electrode pair with the gold-standard or the eye tracker.

\subsubsection*{Saccade Analysis}
Based on the ideal ear electrodes, obtained from the smooth pursuit analysis, we perform further analyses on the saccade data:
(i) We analyse the average saccade signal for each direction and visual angle.
(ii) We analyse the average voltage deflection for each direction and visual angle.
(iii) We predict the visual angle of a saccade from the voltage deflection using a linear regression model.
To perform these tasks, we label the start and end of the \numSaccadesLeftRight saccades based on the gold-standard EOG signals in the direction of the saccade.
In addition, we exclude saccades, for which no clear start and end could be identified.
This leaves us with \numValidSaccades saccades we use for further analysis.

\paragraph{Average Saccade Signal Analysis}
For each of the four directions and visual angles (2.5°, 5°, 7.5°, 10°, 12.5° and 15°), we compute the average signal of the saccades.
This allows us to gain a more comprehensive understanding of the eye movement patterns and the corresponding electrical signals for both the gold-standard EOG and earEOG saccades.

\paragraph{Saccade Amplitudes and Corresponding EOG Voltage Deflections}
In order to understand the relationship between the saccade amplitude and the voltage deflection in the earEOG signals, we calculate the mean voltage deflection for each direction and visual angle.
This further allows us to study the relationship between the saccade amplitude and the voltage deflection in the earEOG signals in comparison to the gold-standard EOG signals.
For each relevant saccade, the voltage deflection is determined by subtracting the mean of the last ten samples of the saccade from the mean of the first ten samples.

\paragraph{Saccade Angle Prediction}
The measurement of voltage deflections through electrooculography (EOG) does not directly provide information on the absolute gaze angle. 
To overcome this limitation, we developed a regression model that predicts the horizontal gaze angle from earEOG or gold-startard EOG data when users were performing saccades as seen in \autoref{fig:saccade-angles}.
The \numValidSaccades valid saccades were all interpolated to contain the same number of samples and were assigned to the respective eye tracker ground truth gaze angle change.
We then computed the voltage deflection by subtracting the mean of the last ten samples in a saccade recording from the mean of the first ten, noting that the actual saccade occurs approximately at the center of the recording.
Using the \textit{scikit-learn} library we trained a linear regression model to predict the change in gaze angle based on the optimal electrode pairing, obtained from the smooth pursuit analysis.
The classification model was evaluated using leave-one-subject out cross-validation and the experiment was carried out for both the earEOG and the EOG signals.

\section*{Results}
In total, 18 participants were recruited for the study.
Because of data corruption issues, two participants had to be excluded from the study.
Therefore, data analysis was conduced using data from \numParticipants participants
(\numParticipantsMale male, \numParticpantsFemale female). The mean age was \meanParticipantAge years (SD = \meanParticipantAgeSD years, range \minParticipantAge to \maxParticipantAge years). Participants were from European (N=\numParticipantsEuropean), Asian (N=\numParticipantsAsian) and Latin-American (N=\numParticipantsLatino) descent.\numParticipantsBeardNo participants indicating no beard, while \numParticipantsBeardYes had a beard, and another \numPariticpansBeardBarely had barely any beard.
Regarding hair length, \numParticipantsHairShort participants had short hair, \numParticipantsHairLong had long hair, and \numParticipantsHairMedium had medium-length hair. 
The mean inside distance from eye to eye was \meanInsideEyeEyeDistance cm (SD = \meanInsideEyeEyeDistanceSD cm). The mean outside distance from eye to eye was \meanOutsideEyeEyeDistance cm (SD = \meanOutsideEyeEyeDistanceSD cm). The mean inside distance from the left eye to the vertical center of the frontal left earEOG electrodes was \DistanceLeftEyeElectrodeCenter cm (SD = \DistanceLeftEyeElectrodeCenterSD cm). The mean distance from the right eye to the vertical center of the frontal right earEOG electrodes was also \DistanceRightEyeElectrodeCenter cm (SD = \DistanceRightEyeElectrodeCenterSD cm).
The following sections present the results and discussion of our evaluation.

\subsection*{Comparison of Electrode Positions}

\begin{figure}
    \centering
    \begin{subfigure}{0.60\textwidth}
        \centering
        \caption{
            Horizontal earEOG montage smooth pursuit correlations in comparison to gold-standard EOG ($\rm r_{EOG}$) and camera-based eye tracking ($\rm r_{CAM}$).}
        \includegraphics[height=6cm]{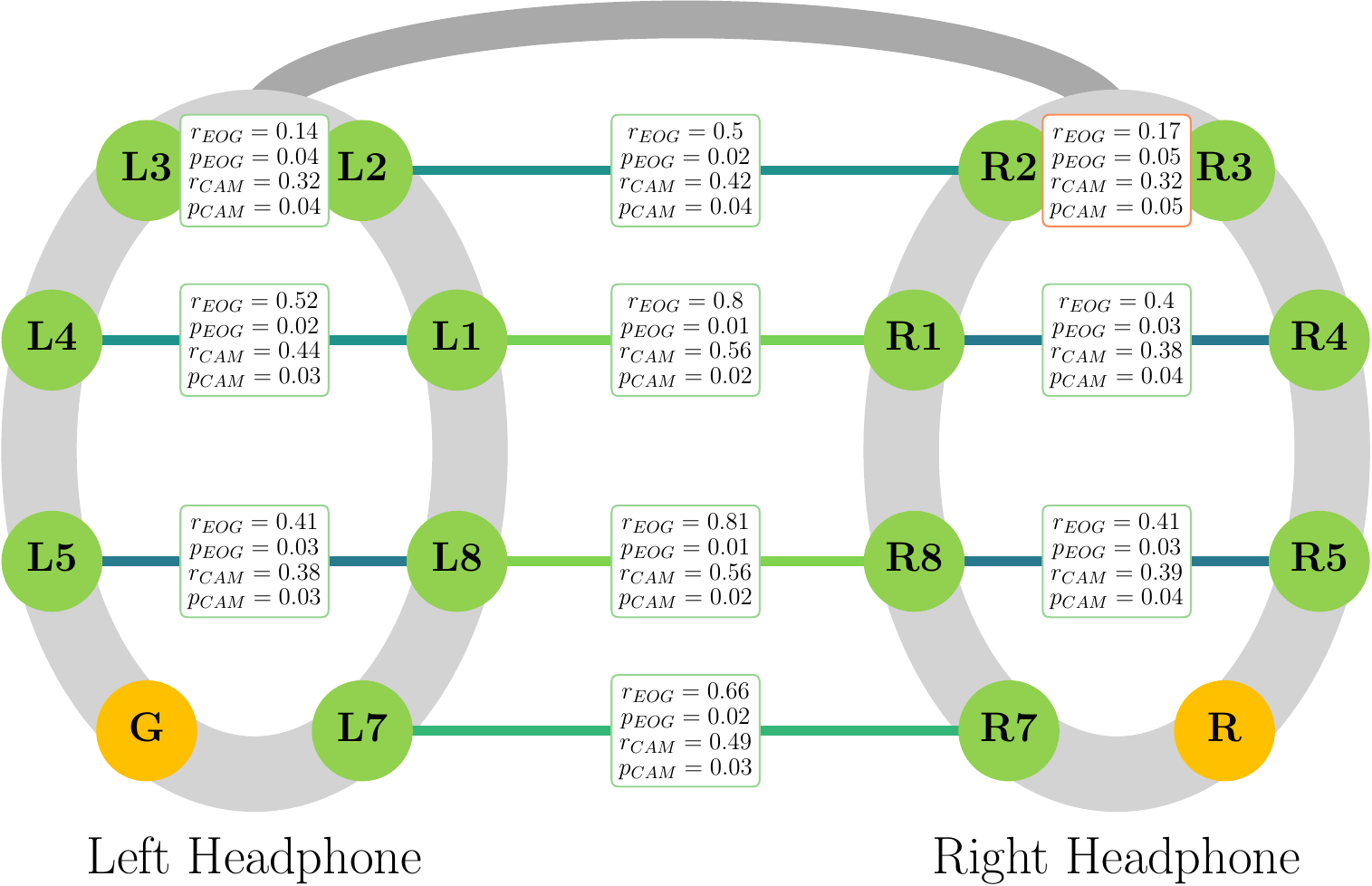}
        \label{fig:corr_horizontal}
    \end{subfigure}
    \hfill
    \begin{subfigure}{0.38\textwidth}
        \centering
        \caption{p-values for pairwise Wilcoxon signed-rank test with respect to $\rm r_{GOLD}$ for horizontal earEOG montages.}
        \includegraphics[height=6cm]{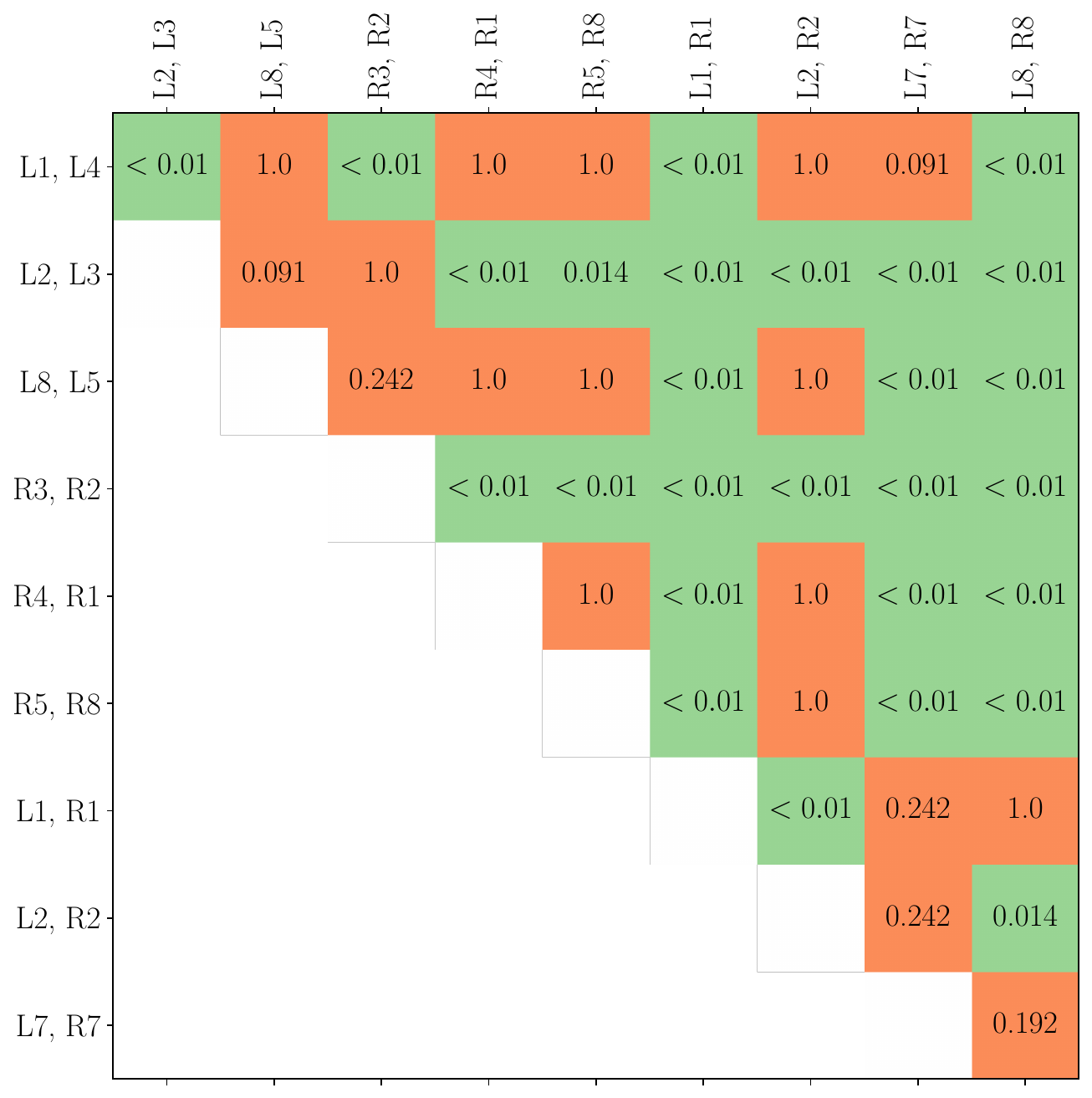}
        \label{fig:corr_horizontal_matrix}
    \end{subfigure}
    \par\bigskip
    \begin{subfigure}{0.60\textwidth}
        \centering
        \caption{Vertical earEOG montage smooth pursuit correlations in comparison to gold-standard EOG ($\rm r_{EOG}$) and camera-based eye tracking ($\rm r_{CAM}$).}
        \includegraphics[height=6cm]{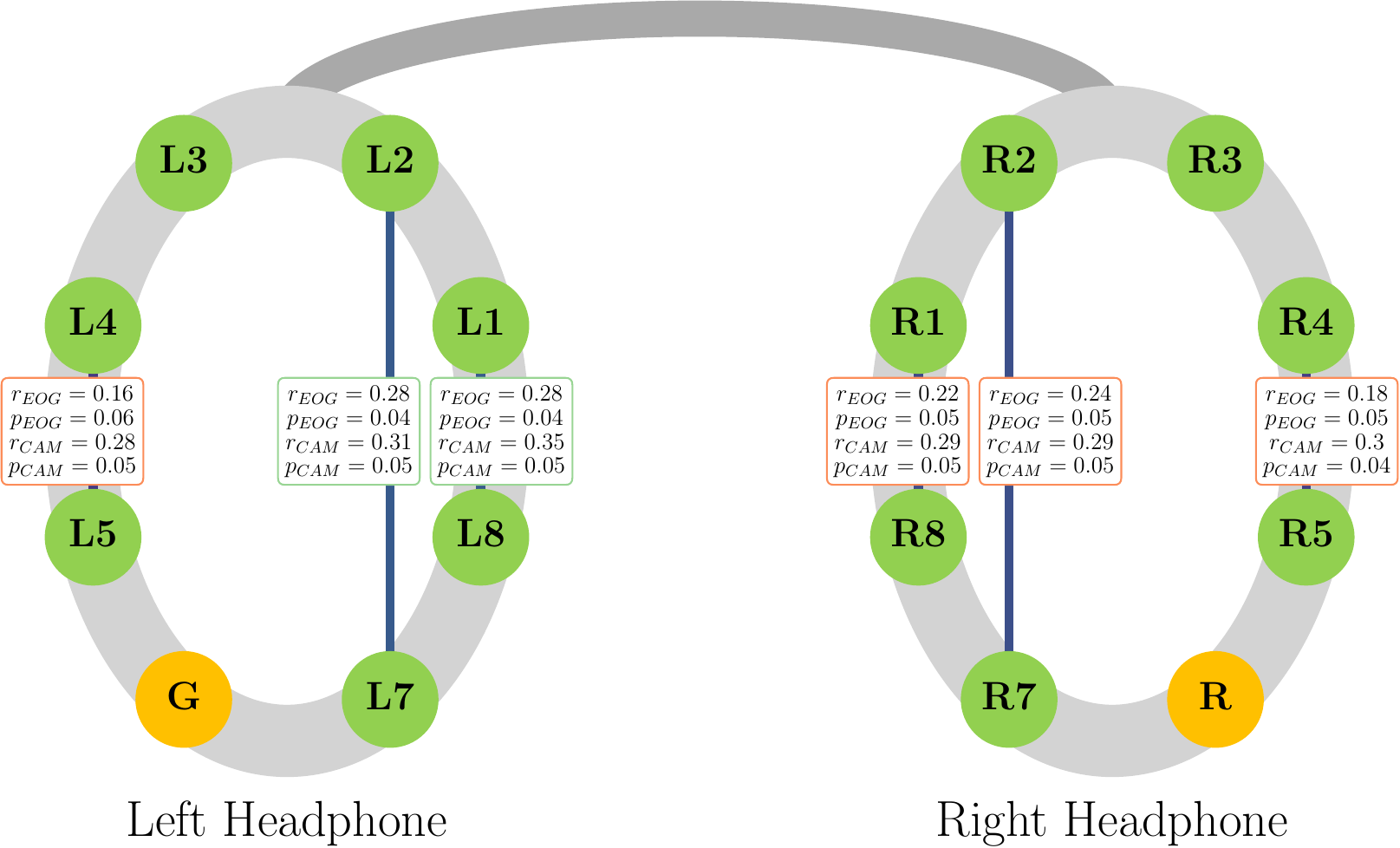}
        \label{fig:corr_vertical}
    \end{subfigure}
    \hfill
    \begin{subfigure}{0.38\textwidth}
        \centering
        \caption{p-values for pairwise Wilcoxon signed-rank test with respect to $\rm r_{CAM}$ for horizontal earEOG montages.}
        \includegraphics[height=6cm]{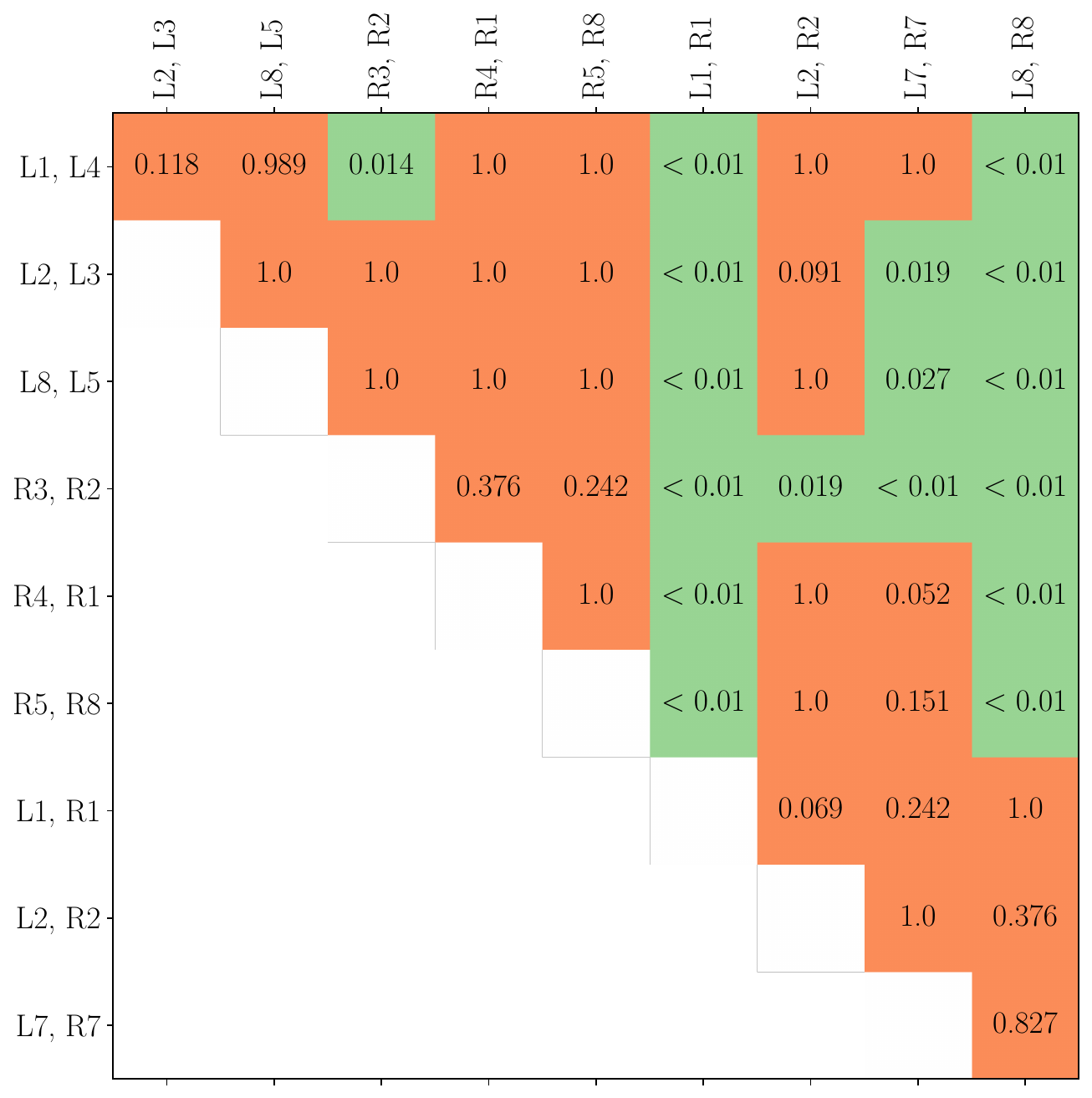}
        \label{fig:corr_horizontal_matrix_camera}
    \end{subfigure}
    
    \caption{
        \textbf{Comparison of earEOG electrode montages to gold-standard EOG and camera-based eye tracking.}
        \ref{fig:corr_horizontal} Correlations of pairwise electrode montages with gold-standard hEOG revealing that electrodes closer to the ears and at eye level yield higher correlations than electrodes on one ear or farther away from eye level;
        \ref{fig:corr_horizontal_matrix} Statistical tests showing p-values for pairwise Wilcoxon signed-rank tests with respect to gold-standard EOG for horizontal montages;
        \ref{fig:corr_vertical} Correlations of pairwise electrode montages with gold-standard vEOG;
        \ref{fig:corr_horizontal_matrix_camera} Statistical tests showing p-values for pairwise Wilcoxon signed-rank tests with respect to the eye tracking data for horizontal montages.
    }
\end{figure}

\paragraph{Horizontal earEOG Electrodes} 
Combining electrodes from the left ear reveals the best electrode pair to be \bestHorizontalCorrleationLeftEarChannelOne-\bestHorizontalCorrleationLeftEarChannelTwo ($r_{\bestHorizontalCorrleationLeftEarChannelOne-\bestHorizontalCorrleationLeftEarChannelTwo}~=\bestHorizontalCorrleationLeftEarEOGRValue$).
On the right ear, the best electrode pair was found to be 
\bestHorizontalCorrleationRightEarChannelOne-\bestHorizontalCorrleationRightEarChannelTwo ($r_{\bestHorizontalCorrleationRightEarChannelOne-\bestHorizontalCorrleationRightEarChannelTwo}~=\bestHorizontalCorrleationRightEarEOGRValue$).
Combining electrodes from both ears increases the correlation further up with the best electrode pair being \bestHorizontalCorrelationChannelOne-\bestHorizontalCorrelationChannelTwo ($\rm r_{\bestHorizontalCorrelationChannelOne-\bestHorizontalCorrelationChannelTwo}~=\bestHorizontalCorrelationEarEogRValue$).
The results are shown in \autoref{fig:corr_horizontal}.
earEOG based on electrodes at eye level have the highest correlation. 
Moving farther away from eye level decreases the correlation.
The smallest correlation to horizontal earEOG is achieved by the montages of L3-L2, R2-R3, and L2-R2, which are the farthest away from the eye level.

The Friedman test with the gold-standard EOG as a ground truth revealed a p-value of \friedmanHorizontal.
As the p-value is below the significance level of 0.05, we can reject the null hypothesis that the electrode montages have the same correlation to the gold-standard EOG.
Therefore, we performed a post-hoc analysis using the Wilcoxon signed-rank test with Bonferroni correction.
The results are shown in \autoref{fig:corr_horizontal_matrix}.

For the camera-based eye tracking ground truth, the Friedman test revealed a p-value of \friedmanHorizontalTobii.
The p-value again is below the significance level of 0.05, and we can reject the null hypothesis that the electrode montages have the same correlation to the camera-based eye tracking.
Again, we performed a post-hoc analysis using the Wilcoxon signed-rank test with Bonferroni correction and the results can be seen in \autoref{fig:corr_horizontal_matrix_camera}.

Notably, the test indicates, that there is no detectable difference between the best horizontal electrode montage (L8-R8) and the second-best electrode pair (L1-R1) in reference to the gold-standard EOG.

\paragraph{Vertical earEOG Electrodes} 
The montages of 
L2-L7, \bestVerticalCorrleationLeftEarChannelOne-\bestVerticalCorrleationLeftEarChannelTwo on the left ear and \bestVerticalCorrleationRightEarChannelOne-\bestVerticalCorrleationRightEarChannelTwo on the right ear yield the strongest correlation (see \autoref{fig:corr_vertical}). Moving farther away from the eyes decreases performance and produces much smaller correlations with \textcolor{red}{L4-L5 and R4-R5} exhibiting the smallest correlation to gold-standard EOG. 

For the vertical earEOG electrodes, the Friedman test yielded a p-value of \friedmanVertical when compared to the gold-standard EOG and a p-value of \friedmanVerticalTobii when compared to camera-based eye tracking.
As both p-value are above the significance level of 0.05, we cannot reject the null hypothesis that the electrode montages have the same correlation to the gold-standard EOG and camera-based eye tracking respectively.
Therefore, no post-hoc analysis was performed.

\paragraph{Discussion} 
Horizontal earEOG electrodes showed promising results.
Signals were generally stronger when measured closer to the eyes, making the combination of electrodes from both ears advantageous as it captures more of the signal.
The electrode pairs placed on the left ear generally exhibit slightly better performance compared to the gold-standard EOG, which may be due to the fact that the gold-standard EOG was recorded from the left eye.

Regarding the vertical earEOG electrodes, we found that the electrodes closest to the ears generally were not the most effective, with the exception of L1-L8.
On both ears, the electrode pairs covering sufficient vertical distance achieve the highest correlation (L2-L7 and R2-R7). 
The performance of vertical earEOG electrodes further diminished as they were placed farther away from the eyes (L4-L5 and R4-R5).
This suggests that a sufficient vertical distance covered by the electrodes on the ears is needed to achieve high correlation.
Overall, vertical earEOG appears to be ineffective for measuring eye movements reliably.

\subsection*{Saccade Analysis}
\begin{figure}[!t]
    \begin{subfigure}[b]{\textwidth}
        \caption{Horizontal saccades}
        \includegraphics[width=\textwidth]{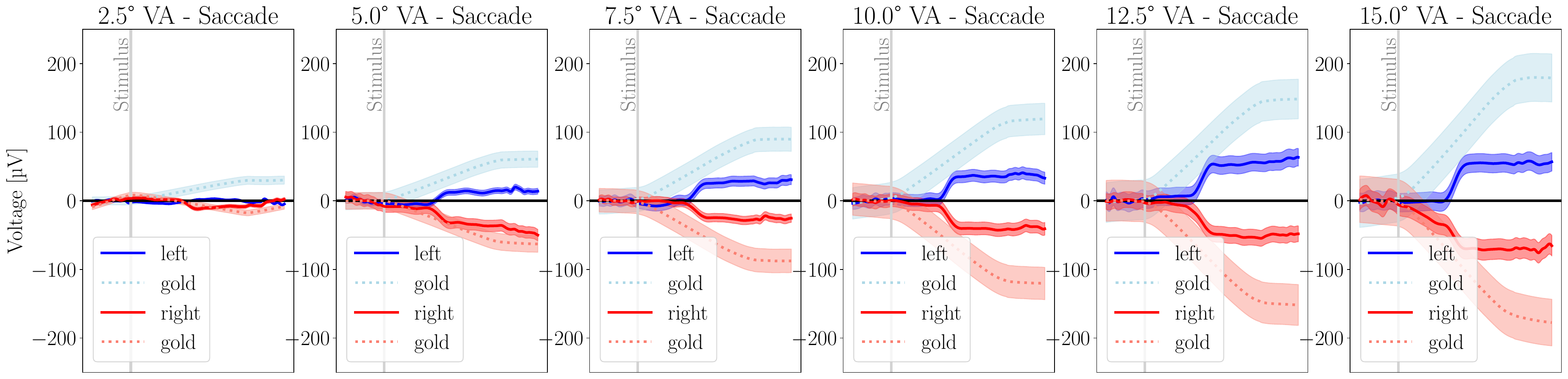}
    \end{subfigure}
    \par\bigskip
    \begin{subfigure}[b]{\textwidth}
        \caption{Vertical saccades}
        \includegraphics[width=\textwidth]{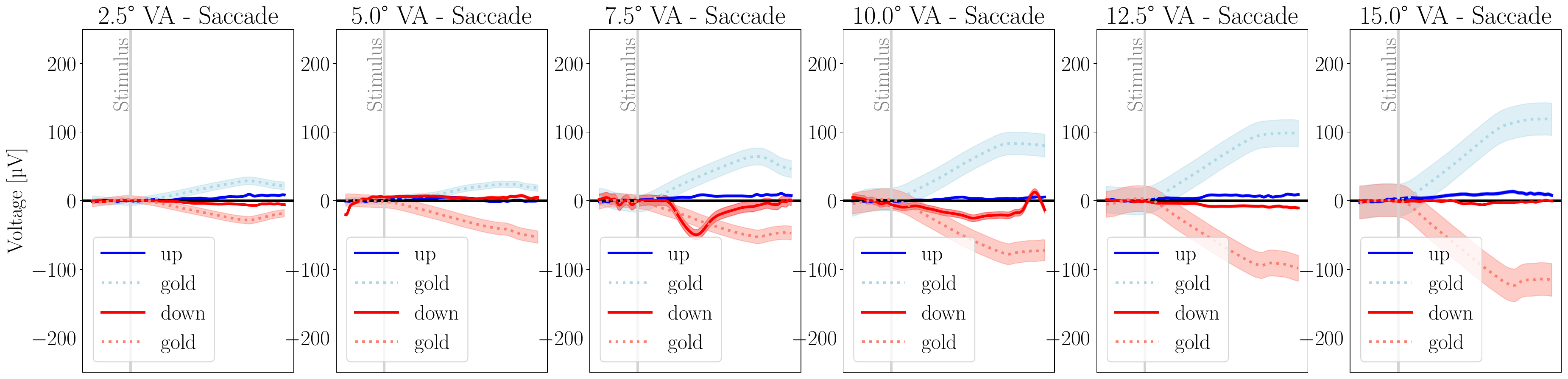}
    \end{subfigure}
    \caption{\textbf{The figure displays the fixation-related EOG signals plotted for different saccade directions (left, right, up, down) and visual angles (2.5°, 5°, 7.5°, 10°, 12.5°, 15°).} The red and blue lines represent the average earEOG signal. The dotted lines represent the average gold-standard EOG signal. The shaded area around the lines represents the standard deviation.}
    \label{fig:va}
\end{figure}

\paragraph{Average Saccade Signal Analysis}
The average saccade analysis is based on the best electrode pair according to the smooth pursuit correlations (\bestHorizontalCorrelationChannelOne-\bestHorizontalCorrelationChannelTwo for horizontal, \bestVerticalCorrelationChannelOne-\bestVerticalCorrelationChannelTwo for vertical saccades).
\autoref{fig:va} provides insight into the average signal of the saccades for each direction and visual angle.
The signals were shifted to zero based on the average voltage difference of the electrode montage before the saccade was performed to better show the relative change per angle and direction.

Upon examining the average saccade signals for each direction and visual angle, we can identify several key features and trends: (i) The EOG signals for saccades in the horizontal direction, irrespective of their amplitude, exhibit a similar waveform shape. This consistency suggests that the EOG system can reliably detect saccades in the horizontal direction and that the waveform shape is characteristic of the saccade direction; (ii) magnitude: The signal magnitude in the horizontal direction generally increases with increasing saccade amplitude, indicating a strong relationship between saccade amplitude and EOG signal magnitude; (iii) The average horizontal saccade signals are reasonably pronounced, allowing for clear identification of the saccades; (iv) vertical saccades exhibit much smaller amplitudes and are less consistent than horizontal earEOG waveforms.

\paragraph{Discussion}
For horizontal saccades, the consistency in signal shape, the relationship between signal duration and amplitude, and the overall signal magnitude trends support the notion that the EOG system near the ears can effectively track eye movements across different amplitudes in the horizontal direction.
As the signal for the vertical saccades is not that consistent or pronounced, this conclusion cannot be made for the vertical saccades.

\subsection*{Saccade Amplitudes and Corresponding EOG Voltage Deflections}
\begin{figure}
    \centering
    \includegraphics[width=\textwidth]{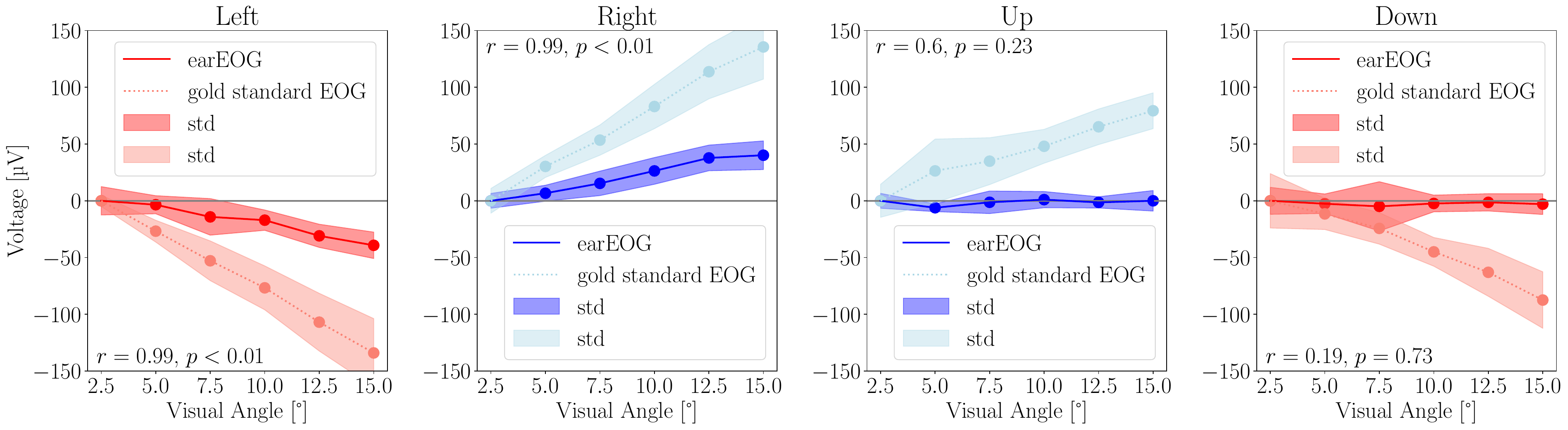}
    \caption{\textbf{Voltage differences for each visual angle and direction of gaze using earEOG and gold-standard EOG}. Scatter plots show the mean voltage differences for each visual angle and direction of gaze using earEOG (red / blue) and gold-standard EOG (coral / lightblue). The filled areas around the lines show the standard deviation.}
    \label{fig:va-comp}
\end{figure}

The average voltage deflections for each visual angle and direction are summarized in \autoref{fig:va-comp}.
In addition, the figure shows the correlation between the absolute voltage deflections at the best earEOG positions 
(\bestHorizontalCorrelationChannelOne-\bestHorizontalCorrelationChannelTwo for horizontal saccades, \bestVerticalCorrelationChannelOne-\bestVerticalCorrelationChannelTwo for vertical saccades) and gold-standard wet electrode EOG.

\paragraph{Horizontal Saccades}
For horizontal saccades, the EOG voltage deflections increased with increasing saccade amplitude and shows a linear trend. The 7.5° left saccade showed a deflection of \saccadeVoltageDeflectionEarEOGleftSevenFive µV, while the 15° left saccade resulted in a deflection of \saccadeVoltageDeflectionEarEOGleftOneFiveZero µV. In the right direction, the 7.5° saccade showed a \saccadeVoltageDeflectionEarEOGrightSevenFive µV deflection and the 15° right saccade resulted in a \saccadeVoltageDeflectionEarEOGrightOneFiveZero µV deflection. This trend suggests that earEOG captures increasing voltage deflections with increasing saccade amplitude.
For horizontal saccades, earEOG and gold-standard EOG deflections were found to be very strongly correlated at $\rm r_{left} = \saccadesleftRValue$, $\rm p_{left} = \saccadesleftPValue$ and $\rm r_{right} = \saccadesrightRValue$, $ \rm p_{right} = \saccadesrightPValue$.

\paragraph{Vertical Saccades}
For vertical saccades, the EOG average voltage deflections did not show a continuous increase with increasing saccade amplitude.
Upward saccades did not result in deflections, with the 7.5° upward saccade showing only a \saccadeVoltageDeflectionEarEOGupSevenFive µV deflection and the 15° upward saccade resulting in only a \saccadeVoltageDeflectionEarEOGupOneFiveZero µV deflection.
Downward saccades also did not result in siginificant deflections, with the 7.5° downward saccade showing only a \saccadeVoltageDeflectionEarEOGdownFiveZero µV deflection and the 15° downward saccade resulting in a \saccadeVoltageDeflectionEarEOGdownOneFiveZero µV deflection.
For vertical saccades, the correlations between earEOG and gold-standard EOG deflections were more varied compared to horizontal saccades. For upward saccades, the correlation was $\rm r_{up} = \saccadesupRValue$. Downward saccades had even weaker correlation of $\rm r_{down} = \saccadesdownRValue$.

\paragraph{Discussion}
The results demonstrate that the earEOG reliably captures horizontal saccades with varying amplitudes but fails to consistently measure voltage changes related to vertical saccades.
The voltage deflections for horizontal saccades show a more consistent, linear trend with increasing amplitude compared to vertical saccades.
The correlation analysis of the absolute voltage deflections between the earEOG and gold-standard EOG provides valuable insights into the earEOG system's performance. The high correlations for horizontal saccades indicate that the earEOG system can reliably track horizontal eye movements.
On the other hand, the more varied correlations for vertical saccades suggest that the earEOG system fails to capture vertical eye movements.

\subsection*{Gaze Angle Prediction}
\begin{figure}
    \centering
    \begin{minipage}[b][][b]{0.35\textwidth}
        \begin{tabular}{l|l|l}
 & earEOG & Gold \\
\midrule
Left 2.5° & 3.54° $\pm$ 6.16° & 0.80° $\pm$ 0.75° \\
Left 5.0° & 2.89° $\pm$ 2.87° & 1.44° $\pm$ 0.88° \\
Left 7.5° & 4.37° $\pm$ 4.34° & 2.51° $\pm$ 1.68° \\
Left 10.0° & 4.21° $\pm$ 3.67° & 3.03° $\pm$ 1.80° \\
Left 12.5° & 4.36° $\pm$ 2.70° & 3.79° $\pm$ 2.45° \\
Left 15.0° & 5.96° $\pm$ 4.14° & 4.52° $\pm$ 2.85° \\
\bf Left \bf Total & \bf 4.24° $\pm$ \bf 4.22° & \bf 2.72° $\pm$ \bf 2.29° \\
\midrule
Right 2.5° & 2.41° $\pm$ 2.03° & 1.46° $\pm$ 1.41° \\
Right 5.0° & 3.18° $\pm$ 2.70° & 1.28° $\pm$ 1.04° \\
Right 7.5° & 4.59° $\pm$ 3.98° & 2.08° $\pm$ 1.30° \\
Right 10.0° & 4.97° $\pm$ 3.79° & 2.95° $\pm$ 1.86° \\
Right 12.5° & 5.11° $\pm$ 4.52° & 3.55° $\pm$ 2.19° \\
Right 15.0° & 6.28° $\pm$ 3.38° & 4.39° $\pm$ 2.67° \\
\bf Right \bf Total & \bf 4.45° $\pm$ \bf 3.74° & \bf 2.64° $\pm$ \bf 2.15° \\
\midrule
\bf Total & \bf 4.34° $\pm$ \bf 3.99° & \bf 2.68° $\pm$ \bf 2.23° \\
            \end{tabular}
            \caption{Mean absolute errors (MAEs) and standard deviations were calculated for both linear regression models: one trained on ear-EOG saccade data and the other on gold-standard EOG saccade data.}
            \label{tab:saccade_linear_regression}
        \end{minipage}
        \hfill
        \begin{minipage}[b][][b]{0.55\textwidth}
            \centering
            \includegraphics[width=\linewidth]{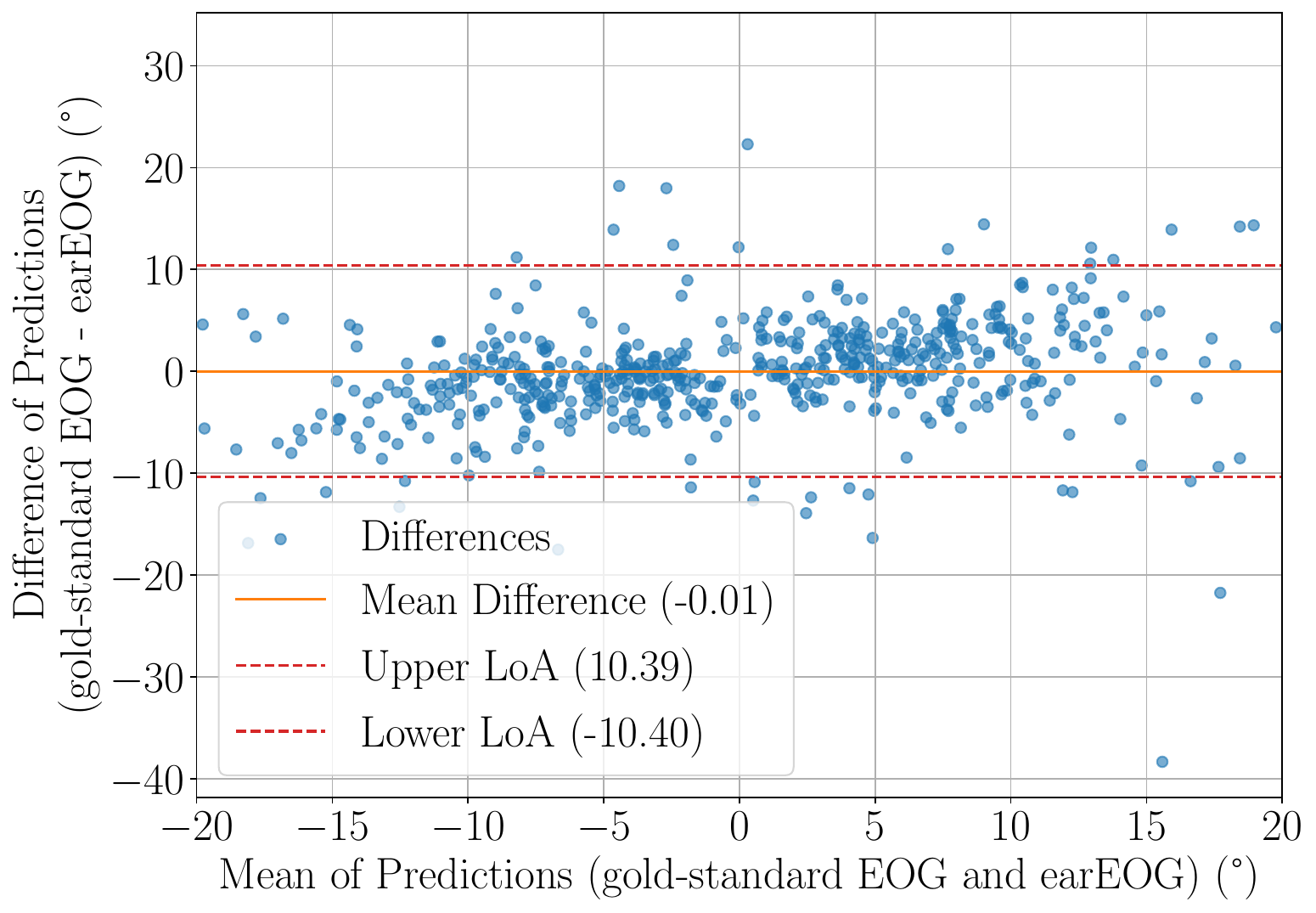}
            \caption{Bland-Altman plot comparing the performance of two linear regression models: one trained on earEOG saccades and the other on gold-standard saccades. \vspace{0.81cm}}
            \label{fig:bland_altman}
        \end{minipage}
\end{figure}

Table \ref{tab:saccade_linear_regression} presents the mean absolute gaze angle errors and standard deviations for the gold-standard and ear-based EOG methods across different gaze angles (2.5°, 5.0°, 7.5°, 10°, 12.5°, and 15°) and directions (left and right). The results for the earEOG are based on the best horizontal electrode pair (\bestHorizontalCorrelationChannelOne-\bestHorizontalCorrelationChannelTwo), as determined in preliminary experiments. As shown in the previous section, the earEOG system cannot reliably capture vertical eye movements; therefore, the up and down directions are excluded from this experiment.
The overall errors for the gold-standard method are smaller than those of the earEOG method in both directions: Left ($\linearRegressionMeanAbsErrorGoldleft^\circ \pm \linearRegressionStdAbsErrorGoldleft^\circ$ vs. $\linearRegressionMeanAbsErrorEarleft^\circ \pm \linearRegressionStdAbsErrorEarleft^\circ$) and right ($\linearRegressionMeanAbsErrorGoldright^\circ \pm \linearRegressionStdAbsErrorGoldright^\circ$ vs. $\linearRegressionMeanAbsErrorEarright^\circ \pm \linearRegressionStdAbsErrorEarright^\circ$).
To evaluate the agreement between the earEOG and the gold-standard EOG, we employ a Bland-Altman plot, shown in \autoref{fig:bland_altman}. The majority of measurements exhibit differences centered around zero, indicating no significant bias.
The mean difference, also displayed on the plot, further supports this observation.
Additionally, the limits of agreement (LoA) are depicted and are rarely exceeded, demonstrating consistent agreement between the two methods.

\paragraph*{Discussion}
The results indicate that the gold-standard method outperforms the ear-based EOG method in terms of mean absolute gaze angle errors across all gaze angles and tested directions.
The overall error for the gold-standard method was almost half of the overall error for the ear-based EOG method, suggesting that the gold-standard method provides more accurate gaze angle measurements.

\section*{Discussion}
In this paper, we investigated the performance of an earEOG system for tracking eye movements, specifically focusing on smooth pursuits and saccades of different directions and amplitudes. Our analysis included the correlations between the earEOG and gold-standard EOG, the examination of average saccade signals and voltage deflections, and absolute gaze angle prediction. The results of our study provide valuable insights into the potential and limitations of the earEOG system for eye movement tracking and analysis.

\subsection*{Vertical vs. Horizontal earEOG}
Our study identified the optimal electrode pairs for measuring vertical and horizontal eye movements as \bestHorizontalCorrelationChannelOne-\bestHorizontalCorrelationChannelTwo for horizontal and \bestVerticalCorrelationChannelOne-\bestVerticalCorrelationChannelTwo for vertical movements.
Horizontal earEOG significantly outperforms vertical earEOG, with much higher correlation values (r = \bestHorizontalCorrelationEarEogRValue, p = \bestHorizontalCorrelationEarEogPValue) for horizontal electrodes compared to vertical ones (r = \bestVerticalCorrelationEarEOGRValue, p = \bestVerticalCorrelationEarEOGPValue) when benchmarked against the gold-standard EOG system.
This is likely due to the fact that horizontal eye movements are more pronounced, stemming from
the larger physical movement of the eyeball and the resulting shift in the electric field of the dipole, making them easier to
measure—even for the gold-standard EOG used in our study.
The consistency in signal shape, duration, and magnitude trends observed in the average saccade signals and the prediction performance further support the earEOG system's ability to effectively track horizontal eye movements.
While the earEOG system reliably tracks horizontal saccades with high correlations and consistent signal patterns, vertical saccades show more variability.
For the smooth pursuit task, the correlation between earEOG and gold-standard EOG were low. This is further highlighted in the
saccade analysis. In the average saccade analysis, the signals are not consistent in comparison to the gold-standard and no linear
trend can be observed in the up and down directions when analyzing the voltage deflections (\autoref{fig:va-comp}).
This may stem from anatomical factors, as vertical eye movements produce smaller electrical signals, even for the gold-standard measurement positions as seen in \autoref{fig:va-comp}.
These findings demonstrate the feasibility of horizontal earEOG for practical applications, whereas vertical earEOG appears to be impractical. 
Identified electrode pairs can be used in future research and clinical settings conducting similar research wit hear-based EOG.

which indicates bad performance in practical applications

\subsection*{Comparison to Related Work}
Favre-Félix et al.\cite{favre2019absolute} investigated 30\textdegree~horizontal saccades using in-ear electrodes. They reported a voltage change of approx. 50 $\rm\mu\text{V}$ during saccades.
While amplitude comparisons are influenced by factors like noise floor, gain or skin preparation, and direct comparisons without the same measurement equipment are difficult, earEOG based on our headphone setup achieves up to approx. 150 $\rm\mu\text{V}$, suggesting that saccades can be measured more reliably using the periauricular positioning of electrodes.

Manabe et al.\cite{manabe2006full} used a related headphone setup with gaze targets that were 20\textdegree~apart.
The electrodes where arranged around the ear, with four electrodes on each side.
They reported an overall absolute gaze angle estimation error of 4.4° horizontal, which closely aligns with our findings, and 8.3° vertical in a study with six participants.
Similar to our finding, the vertical error exceeded the horizontal error. They calibrate the model per user which increases the performance compared to our general model which is implemented as a one-fits-all approach.

Barbara et al. \cite{barbaraInterfacingSpellerUsing2016} explored the use of EOG glasses to measure saccades for human-computer interaction.  
Their study demonstrated a saccade detection accuracy of 73.38\% using a threshold-based classification algorithm.  
Notably, they distinguished between horizontal and vertical saccades using subject-specific thresholds during calibration. 
In our work, we did not perform saccade detection and instead, predict the angle of saccades.

\subsection*{Limitations}
Our study extends prior research on earEOG technology, offering new insights through a laboratory investigation with 16 participants.
While this lab-study provides insight into the applicability of earEOG systems, some improvements could be made to further showcase its feasibility in the real world.
Firstly, the paper assumes a center resting position for gaze, and therefore does not investigate any relative saccadic movements. As saccades play a crucial role in visual perception, this assumption may limit the generalizability of our findings.
Secondly, the study does not account for head movements during saccades, assuming instead that the head is fixed in space. However, turning the head is an integral part of gaze, and this oversight may introduce bias in our results.
Lastly, our gaze angle prediction method assumes that saccades have already been detected, and only predicts the angle of gaze. In a real-world system, the isolation of saccades would be a necessary step.

\section*{Conclusion}

In this work, we demonstrated the potential of ear-based EOG (earEOG) for measuring eye movements with varying amplitudes and directions. Our results establish the feasibility of tracking horizontal eye movements using the earEOG system, with the optimal horizontal electrode pair identified as \bestHorizontalCorrelationChannelOne-\bestHorizontalCorrelationChannelTwo ($\rm r_{EOG} = \bestHorizontalCorrelationEarEogRValue, p=\bestHorizontalCorrelationEarEogPValue$; $\rm r_{CAM} = \bestHorizontalCorrelationCamRValue, p=\bestHorizontalCorrelationCamPValue$). This pair exhibited a linear relationship between the visual angle of a saccade and the corresponding voltage deflection ($\rm r_{left} = \saccadesleftRValue$, $\rm p_{left} = \saccadesleftPValue$ and $\rm r_{right} = \saccadesrightRValue$, $ \rm p_{right} = \saccadesrightPValue$), showcasing its potential for precise tracking.

Although the gold-standard EOG method outperforms earEOG in terms of gaze angle prediction accuracy, earEOG offers significant advantages for unobtrusive, day-to-day use.
For instance, our system could pave the way for applications such as detecting dizziness or other vestibular disorders in real-world settings. However, the results indicate that earEOG is not suitable for measuring vertical eye movements.
The best vertical electrode pair showed only weak correlation when users performed smoot pursuits with the best pair being \bestVerticalCorrelationChannelOne-\bestVerticalCorrelationChannelTwo ($\rm r_{EOG} = \bestVerticalCorrelationEarEOGRValue, p=\bestVerticalCorrelationEarEOGPValue$; $\rm r_{CAM} = \bestVerticalCorrelationCamRValue, p=\bestVerticalCorrelationCamPValue$).
Voltage deflections also were significantly less pronounced then the gold-standard EOG ($\rm r_{up} = \saccadesupRValue$, $\rm p_{up} = \saccadesupPValue$ and $\rm r_{down} = \saccadesdownRValue$, $ \rm p_{down} = \saccadesdownPValue$)

In summary, earEOG represents a promising approach for measuring horizontal eye movements, offering potential for practical applications in clinical diagnostics (e.g., dizziness detection) and gaze-based human-computer interaction.
Future work should focus on enhancing system design and signal processing to further improve accuracy and explore its broader applicability.

\section*{Data and Code Availability}
The collected data and code are available from the corresponding author upon reasonable request.

\bibliography{sample}

\begin{thebibliography}{10}
\urlstyle{rm}
\expandafter\ifx\csname url\endcsname\relax
  \def\url#1{\texttt{#1}}\fi
\expandafter\ifx\csname urlprefix\endcsname\relax\def\urlprefix{URL }\fi
\expandafter\ifx\csname doiprefix\endcsname\relax\def\doiprefix{DOI: }\fi
\providecommand{\bibinfo}[2]{#2}
\providecommand{\eprint}[2][]{\url{#2}}

\bibitem{gibaldi2017evaluation}
\bibinfo{author}{Gibaldi, A.}, \bibinfo{author}{Vanegas, M.},
  \bibinfo{author}{Bex, P.~J.} \& \bibinfo{author}{Maiello, G.}
\newblock \bibinfo{journal}{\bibinfo{title}{Evaluation of the tobii eyex eye
  tracking controller and matlab toolkit for research}}.
\newblock {\emph{\JournalTitle{Behavior research methods}}}
  \textbf{\bibinfo{volume}{49}}, \bibinfo{pages}{923--946}
  (\bibinfo{year}{2017}).

\bibitem{chennamma2013survey}
\bibinfo{author}{Chennamma, H.} \& \bibinfo{author}{Yuan, X.}
\newblock \bibinfo{journal}{\bibinfo{title}{A survey on eye-gaze tracking
  techniques}}.
\newblock {\emph{\JournalTitle{arXiv preprint arXiv:1312.6410}}}
  (\bibinfo{year}{2013}).

\bibitem{dhuliawala2016smooth}
\bibinfo{author}{Dhuliawala, M.} \emph{et~al.}
\newblock \bibinfo{title}{Smooth eye movement interaction using eog glasses}.
\newblock In \emph{\bibinfo{booktitle}{Proceedings of the 18th ACM
  International Conference on Multimodal Interaction}},
  \bibinfo{pages}{307--311} (\bibinfo{year}{2016}).

\bibitem{bulling2010eye}
\bibinfo{author}{Bulling, A.}, \bibinfo{author}{Ward, J.~A.},
  \bibinfo{author}{Gellersen, H.} \& \bibinfo{author}{Tr{\"o}ster, G.}
\newblock \bibinfo{journal}{\bibinfo{title}{Eye movement analysis for activity
  recognition using electrooculography}}.
\newblock {\emph{\JournalTitle{IEEE transactions on pattern analysis and
  machine intelligence}}} \textbf{\bibinfo{volume}{33}},
  \bibinfo{pages}{741--753} (\bibinfo{year}{2010}).

\bibitem{zhu2014eog}
\bibinfo{author}{Zhu, X.} \emph{et~al.}
\newblock \bibinfo{title}{Eog-based drowsiness detection using convolutional
  neural networks}.
\newblock In \emph{\bibinfo{booktitle}{2014 International Joint Conference on
  Neural Networks (IJCNN)}}, \bibinfo{pages}{128--134}
  (\bibinfo{organization}{IEEE}, \bibinfo{year}{2014}).

\bibitem{goebel1992prevalence}
\bibinfo{author}{Goebel, J.~A.} \& \bibinfo{author}{Garcia, P.}
\newblock \bibinfo{journal}{\bibinfo{title}{Prevalence of post-headshake
  nystagmus in patients with caloric deficits and vertigo}}.
\newblock {\emph{\JournalTitle{Otolaryngology--Head and Neck Surgery}}}
  \textbf{\bibinfo{volume}{106}}, \bibinfo{pages}{121--127}
  (\bibinfo{year}{1992}).

\bibitem{lee2016real}
\bibinfo{author}{Lee, K.-R.}, \bibinfo{author}{Chang, W.-D.},
  \bibinfo{author}{Kim, S.} \& \bibinfo{author}{Im, C.-H.}
\newblock \bibinfo{journal}{\bibinfo{title}{Real-time “eye-writing”
  recognition using electrooculogram}}.
\newblock {\emph{\JournalTitle{IEEE Transactions on Neural Systems and
  Rehabilitation Engineering}}} \textbf{\bibinfo{volume}{25}},
  \bibinfo{pages}{37--48} (\bibinfo{year}{2016}).

\bibitem{10.1145/3715669.3723110}
\bibinfo{author}{Hansen, A.} \emph{et~al.}
\newblock \bibinfo{title}{Bodypursuits: Exploring smooth pursuit gaze
  interaction based on body motion targets}.
\newblock In \emph{\bibinfo{booktitle}{Proceedings of the 2025 Symposium on Eye
  Tracking Research and Applications}}, ETRA '25,
  \doiprefix\url{10.1145/3715669.3723110} (\bibinfo{publisher}{Association for
  Computing Machinery}, \bibinfo{address}{New York, NY, USA},
  \bibinfo{year}{2025}).

\bibitem{bulling2008robust}
\bibinfo{author}{Bulling, A.}, \bibinfo{author}{Ward, J.~A.},
  \bibinfo{author}{Gellersen, H.} \& \bibinfo{author}{Tr{\"o}ster, G.}
\newblock \bibinfo{title}{Robust recognition of reading activity in transit
  using wearable electrooculography}.
\newblock In \emph{\bibinfo{booktitle}{Pervasive Computing: 6th International
  Conference, Pervasive 2008 Sydney, Australia, May 19-22, 2008 Proceedings
  6}}, \bibinfo{pages}{19--37} (\bibinfo{organization}{Springer},
  \bibinfo{year}{2008}).

\bibitem{kunze2015much}
\bibinfo{author}{Kunze, K.}, \bibinfo{author}{Katsutoshi, M.},
  \bibinfo{author}{Uema, Y.} \& \bibinfo{author}{Inami, M.}
\newblock \bibinfo{title}{How much do you read? counting the number of words a
  user reads using electrooculography}.
\newblock In \emph{\bibinfo{booktitle}{Proceedings of the 6th Augmented Human
  International Conference}}, \bibinfo{pages}{125--128} (\bibinfo{year}{2015}).

\bibitem{favre2017steering}
\bibinfo{author}{Favre-Felix, A.}, \bibinfo{author}{Hietkamp, R.},
  \bibinfo{author}{Graversen, C.}, \bibinfo{author}{Dau, T.} \&
  \bibinfo{author}{Lunner, T.}
\newblock \bibinfo{title}{Steering of audio input in hearing aids by eye gaze
  through electrooculography}.
\newblock In \emph{\bibinfo{booktitle}{Proceedings of the International
  Symposium on Auditory and Audiological Research}}, vol.~\bibinfo{volume}{6},
  \bibinfo{pages}{135--142} (\bibinfo{year}{2017}).

\bibitem{favre2017real}
\bibinfo{author}{Favre-F{\'e}lix, A.}, \bibinfo{author}{Graversen, C.},
  \bibinfo{author}{Dau, T.} \& \bibinfo{author}{Lunner, T.}
\newblock \bibinfo{title}{Real-time estimation of eye gaze by in-ear
  electrodes}.
\newblock In \emph{\bibinfo{booktitle}{2017 39th Annual International
  Conference of the IEEE Engineering in Medicine and Biology Society (EMBC)}},
  \bibinfo{pages}{4086--4089} (\bibinfo{organization}{IEEE},
  \bibinfo{year}{2017}).

\bibitem{favre2019absolute}
\bibinfo{author}{Favre-F{\'e}lix, A.} \emph{et~al.}
\newblock \bibinfo{journal}{\bibinfo{title}{Absolute eye gaze estimation with
  biosensors in hearing aids}}.
\newblock {\emph{\JournalTitle{Frontiers in neuroscience}}}
  \textbf{\bibinfo{volume}{13}}, \bibinfo{pages}{1294} (\bibinfo{year}{2019}).

\bibitem{hladek2018real}
\bibinfo{author}{Hl{\'a}dek, L.}, \bibinfo{author}{Porr, B.} \&
  \bibinfo{author}{Brimijoin, W.~O.}
\newblock \bibinfo{journal}{\bibinfo{title}{Real-time estimation of horizontal
  gaze angle by saccade integration using in-ear electrooculography}}.
\newblock {\emph{\JournalTitle{Plos one}}} \textbf{\bibinfo{volume}{13}},
  \bibinfo{pages}{e0190420} (\bibinfo{year}{2018}).

\bibitem{manabe2013conductive}
\bibinfo{author}{Manabe, H.}, \bibinfo{author}{Fukumoto, M.} \&
  \bibinfo{author}{Yagi, T.}
\newblock \bibinfo{title}{Conductive rubber electrodes for earphone-based eye
  gesture input interface}.
\newblock In \emph{\bibinfo{booktitle}{Proceedings of the 2013 International
  Symposium on Wearable Computers}}, \bibinfo{pages}{33--40}
  (\bibinfo{year}{2013}).

\bibitem{manabe2006full}
\bibinfo{author}{Manabe, H.} \& \bibinfo{author}{Fukumoto, M.}
\newblock \bibinfo{title}{Full-time wearable headphone-type gaze detector}.
\newblock In \emph{\bibinfo{booktitle}{CHI'06 Extended Abstracts on Human
  Factors in Computing Systems}}, \bibinfo{pages}{1073--1078}
  (\bibinfo{year}{2006}).

\bibitem{vidal2013pursuits}
\bibinfo{author}{Vidal, M.}, \bibinfo{author}{Bulling, A.} \&
  \bibinfo{author}{Gellersen, H.}
\newblock \bibinfo{title}{Pursuits: spontaneous interaction with displays based
  on smooth pursuit eye movement and moving targets}.
\newblock In \emph{\bibinfo{booktitle}{Proceedings of the 2013 ACM
  international joint conference on Pervasive and ubiquitous computing}},
  \bibinfo{pages}{439--448} (\bibinfo{year}{2013}).

\bibitem{robinson1965mechanics}
\bibinfo{author}{Robinson, D.~A.}
\newblock \bibinfo{journal}{\bibinfo{title}{The mechanics of human smooth
  pursuit eye movement.}}
\newblock {\emph{\JournalTitle{The Journal of Physiology}}}
  \textbf{\bibinfo{volume}{180}}, \bibinfo{pages}{569} (\bibinfo{year}{1965}).

\bibitem{OpencEEG55:online}
\bibinfo{author}{OpenBCI}.
\newblock \bibinfo{title}{Open-ceegrid kit – openbci online store}.
\newblock \bibinfo{note}{[Online; accessed 2025-01-13]}.

\bibitem{awsmabdullahReviewFilteringTechniques2023}
\bibinfo{author}{{Aws M Abdullah}}, \bibinfo{author}{{Ali R Ibrahim}},
  \bibinfo{author}{{Ammar A Al-Hamadani}}, \bibinfo{author}{{Mohammed K
  Al-Obaidi}} \& \bibinfo{author}{{Anas F Ahmed}}.
\newblock \bibinfo{journal}{\bibinfo{title}{A review for filtering techniques
  of the {{Electrooculography}} ({{EOG}}) signals}}.
\newblock {\emph{\JournalTitle{Global Journal of Engineering and Technology
  Advances}}} \textbf{\bibinfo{volume}{16}}, \bibinfo{pages}{163--171}
  (\bibinfo{year}{2023}).

\bibitem{s23062944}
\bibinfo{author}{Murugan, S.}, \bibinfo{author}{Sivakumar, P.~K.},
  \bibinfo{author}{Kavitha, C.}, \bibinfo{author}{Harichandran, A.} \&
  \bibinfo{author}{Lai, W.-C.}
\newblock \bibinfo{journal}{\bibinfo{title}{An electro-oculogram (eog) sensor's
  ability to detect driver hypovigilance using machine learning}}.
\newblock {\emph{\JournalTitle{Sensors}}} \textbf{\bibinfo{volume}{23}}
  (\bibinfo{year}{2023}).

\bibitem{barbaraInterfacingSpellerUsing2016}
\bibinfo{author}{Barbara, N.} \& \bibinfo{author}{Camilleri, T.~A.}
\newblock \bibinfo{title}{Interfacing with a speller using {{EOG}} glasses}.
\newblock In \emph{\bibinfo{booktitle}{2016 {{IEEE International Conference}}
  on {{Systems}}, {{Man}}, and {{Cybernetics}} ({{SMC}})}},
  \bibinfo{pages}{001069--001074} (\bibinfo{year}{2016}).

\end{thebibliography}

\section*{Acknowledgements}
This work was partially supported through the KD$^2$School and the KD$^2$Lab at the Karlsruhe Institute of Technology (KIT) which is funded by the German Research Foundation (DFG) and by the Carl-Zeiss-Stiftung (Carl-Zeiss-Foundation) as part of the project "JuBot - Staying young with robots". We thank Kathrin Blum for her preliminary experiments based on cEEGrid electrodes.

\section*{Author contributions statement}

All authors conceived the study protocol and wrote as well as reviewed the manuscript. 
T.K. and T.R. contributed the initial concept and implemented the study software.
M.K. assembled the study apparatus and collected the data.
T.K., T.R., M.K., and C.C. analyzed the data.

\section*{Competing Interests}
The authors declare no competing interests.

\end{document}